\documentclass[usenatbib]{mn2e}

\usepackage{psfig,morefloats,url}
\usepackage{graphicx}
\usepackage{natbib}

\newcommand{\degrees}{\ifmmode^{\circ}\else$^{\circ}$\fi}
\newcommand{\amin}{\ifmmode^{\prime}\else$^{\prime}$\fi}
\newcommand{\asec}{\ifmmode^{\prime\prime}\else$^{\prime\prime}$\fi}
\newcommand{\psr}{PSR~J1903$+$0327}
\newcommand{\binary}{J1903$+$0327 binary system }

\title[\psr]{On the nature and evolution of the unique binary pulsar J1903+0327} 
\author[Freire et al.]
{P.~C.~C. Freire$^{1}$, C. G. Bassa$^{2}$, N. Wex$^{1}$, 
I.~H. Stairs$^{3}$, D.~J. Champion$^{1}$,
\newauthor
S.~M. Ransom$^{4}$, P. Lazarus$^{5}$, V.~M. Kaspi$^{5}$,
J.~W.~T. Hessels$^{6,7}$, M. Kramer$^{1}$,
\newauthor
J.~M. Cordes$^{8}$, J.~P.~W. Verbiest$^{1,9}$
P. Podsiadlowski$^{10}$, D.~J. Nice$^{11}$, J.~S. Deneva$^{12}$,  
\newauthor
D.~R. Lorimer$^{9,13}$,
B.~W. Stappers$^{2}$, M.~A. McLaughlin$^{9,13}$ and F. Camilo$^{14}$
\\
$^{1}$ Max-Planck-Institut f\"ur Radioastronomie, Auf dem H\"ugel 69, D-53121 Bonn, Germany\\
$^{2}$ Jodrell Bank Centre for Astrophysics, Univ. of Manchester, Oxford Rd, Manchester, M13 9PL UK\\
$^{3}$ Dept.~of Physics and Astronomy, Univ.~of British Columbia, 6224 Agricultural Rd., Vancouver, BC V6T 1Z1, Canada\\
$^{4}$ National Radio Astronomy Observatory, 520 Edgemont Rd., Charlottesville, VA 22903, USA\\
$^{5}$ Dept.~of Physics, McGill Univ., Montreal, QC H3A 2T8, Canada\\
$^{6}$ Netherlands Institute for Radio Astronomy (ASTRON), Postbus 2, 7990 AA Dwingeloo, The Netherlands\\
$^{7}$ Astronomical Institute ``Anton Pannekoek,'' Univ. of Amsterdam, 1098 SJ Amsterdam, The Netherlands\\
$^{8}$ Dept. of Astronomy, Cornell Univ., Ithaca, NY 14853, USA\\
$^{9}$ Dept. of Physics, West Virginia Univ., Hodges Hall, Morgantown, WV 26506-6315, USA\\
$^{10}$ Dept. of Astronomy, Oxford Univ., Oxford OX1 3RH, UK\\
$^{11}$ Dept. of Physics, Lafayette College, Easton, PA 18042, USA\\
$^{12}$ Arecibo Observatory, HC 3 53995, Arecibo, PR 00612, USA\\
$^{13}$ National Radio Astronomy Observatory, Green Bank, WV 24944, USA\\
$^{14}$ Columbia Astrophysics Laboratory, Columbia Univ., 550 West 120th St., New York, NY 10027, USA}
\date{Re-submitted version}
\begin{document}
\maketitle

\begin{abstract} 
 \psr, a millisecond pulsar in an eccentric ($e = 0.44$) 95-day orbit
 with a ($\sim\,1\,M_{\odot}$) companion poses a challenge to our
 understanding of stellar evolution in binary and multiple-star systems.
 Here we describe optical and radio observations which rule out most of the
 scenarios proposed to explain formation of this system.
 Radio timing measurements of three post-Keplerian effects
 yield the most precise measurement of the
 mass of a millisecond pulsar to date: $1.667 \pm 0.021$ solar masses
 (99.7\% confidence limit). This rules out some equations of state for
 super-dense matter, furthermore it is consistent with spin-up of the
 pulsar by mass accretion, as suggested by its short spin period and
 low magnetic field. Optical spectroscopy of a proposed main sequence
 counterpart show that its orbital motion mirrors the pulsar's
 95-day orbit; being therefore its binary companion. This finding rules
 out a previously suggested scenario which proposes that the system is
 presently a hierarchical triple. Conventional binary
 evolution scenarios predict that, after recycling a neutron star into
 a millisecond pulsar, the binary companion should become a white dwarf
 and its orbit should be nearly circular. This suggests that
 if \psr\,was recycled, its present companion was not responsible for it.
 The optical detection also provides a measurement of the systemic radial
 velocity of the binary; this and the proper motion measured from pulsar
 timing allow the determination of the systemic 3-D velocity in the Galaxy.
 We find that the system is always within 270 pc of the plane of
 the Galaxy, but always more than 3 kpc away from the Galactic centre.
 Thus an exchange
 interaction in a dense stellar environment (like a globular cluster or the
 Galactic centre) is not likely to be the origin of this system.
 We suggest that after the supernova that formed it, the neutron star
was in a tight orbit with a main-sequence star, the present
companion was a tertiary farther out. The neutron star then accreted
matter from its evolving inner MS companion, forming a millisecond pulsar.
The former donor star then disappears, either due to a chaotic
3-body interaction with the outer star (caused by the expansion of the
inner orbit that necessarily results from mass transfer), or in the case of
a very compact inner system, due to ablation/accretion by the newly formed
millisecond pulsar. We discuss in detail the possible evolution of
such a system before the supernova.
\end{abstract}


\section{INTRODUCTION}\label{sec:intro}
\psr\, was the first millisecond pulsar (MSP\footnote{We define a
millisecond pulsar as a pulsar with spin period $P \leq 20$~ms and with
a low surface magnetic field, $B\sim 10^{8-9}$\,G.}) discovered in the
ongoing Arecibo L-band Feed Array (ALFA) pulsar survey \cite{cfl+06}. In the
discovery paper~\cite{crl+08}, we presented the results of phase-coherent
radio timing of this pulsar carried out with the Green Bank and Arecibo
radio telescopes. These quickly revealed that the pulsar was in a 95-day
orbit around a 1 solar mass ($M_{\odot}$) companion. This object is
remarkable for being the first (and thus far, the only) disk MSP
known to have an eccentric ($e = 0.44$) orbit. In globular clusters (GCs)
there are several binary MSPs with eccentric orbits; but those are thought
to be caused by perturbations of the binary systems by occasional
close interactions with other stars.

Coincident with the pulsar position derived from the timing, a
star was found whose near-infrared magnitudes were consistent with a
$1\,M_{\odot}$ main-sequence star at the distance and reddening estimated for
PSR J1903+0327. It was not known then whether this was just an unlikely
($\sim$ 2.6\%) chance alignment or whether the star is genuinely
associated with \psr, and if so whether it is the binary companion
responsible for the 95-day orbit of the pulsar. Such a finding would
be surprising, as the conventional understanding of MSP
evolution posits that such a neutron star (NS) is spun up to high spin
frequencies by accretion of matter and angular momentum from a companion
star while the companion passes through a giant phase \cite{bv91};
this circularises the system, and a recycled MSP is left
orbiting a low-mass white dwarf (the remnant core of the donor) in a
low-eccentricity orbit ($e < 10^{-3}$; Phinney 1992\nocite{phi92}).
Until the discovery of \psr\, all known MSPs in the Galactic disk
had such low-eccentricity orbits.
For reviews, see Phinney~\&~Kulkarni~(1994)\nocite{pk94},
Stairs~(2004)\nocite{sta04}, Tauris \& van den Heuvel (2006)\nocite{th06}.

For these reasons, Champion et al. (2008)\nocite{crl+08}
proposed that \psr\, may be part of a triple system where the 95-day orbit
of the pulsar is caused by a massive unseen WD and the third member is the
star detected in the near-infrared. The latter is in a long-period
orbit and drives the eccentricity of the inner pair through the Kozai mechanism
\cite{koz62}. An alternative possibility, also discussed in \cite{crl+08} is
that the companion to \psr\, in the 95-day orbit is the star detected in
the near-infrared, but that this eccentric, unusual system originated in
an exchange interaction in a dense stellar environment, like a globular
cluster.

In this paper, we present new optical measurements and further radio timing
of \psr\, obtained with the aim of testing these scenarios.
The plan for the rest of this paper is as follows. The optical
and radio observations are described \S~\ref{sec:observations}.
The immediate results from these observations are described in
\S~\ref{sec:results}. In \S~\ref{sec:discussion} we discuss the
implications of these results regarding the formation and
evolution of this system. In \S~\ref{sec:formation} we
discuss how this system might have formed. We summarise our main
conclusions in \S~\ref{sec:conclusions}.

\section{Observations}
\label{sec:observations}

\subsection{Optical observations}
\label{sec:optical}

\begin{figure}
  \label{fig:finder}
  \includegraphics[width=8.2cm]{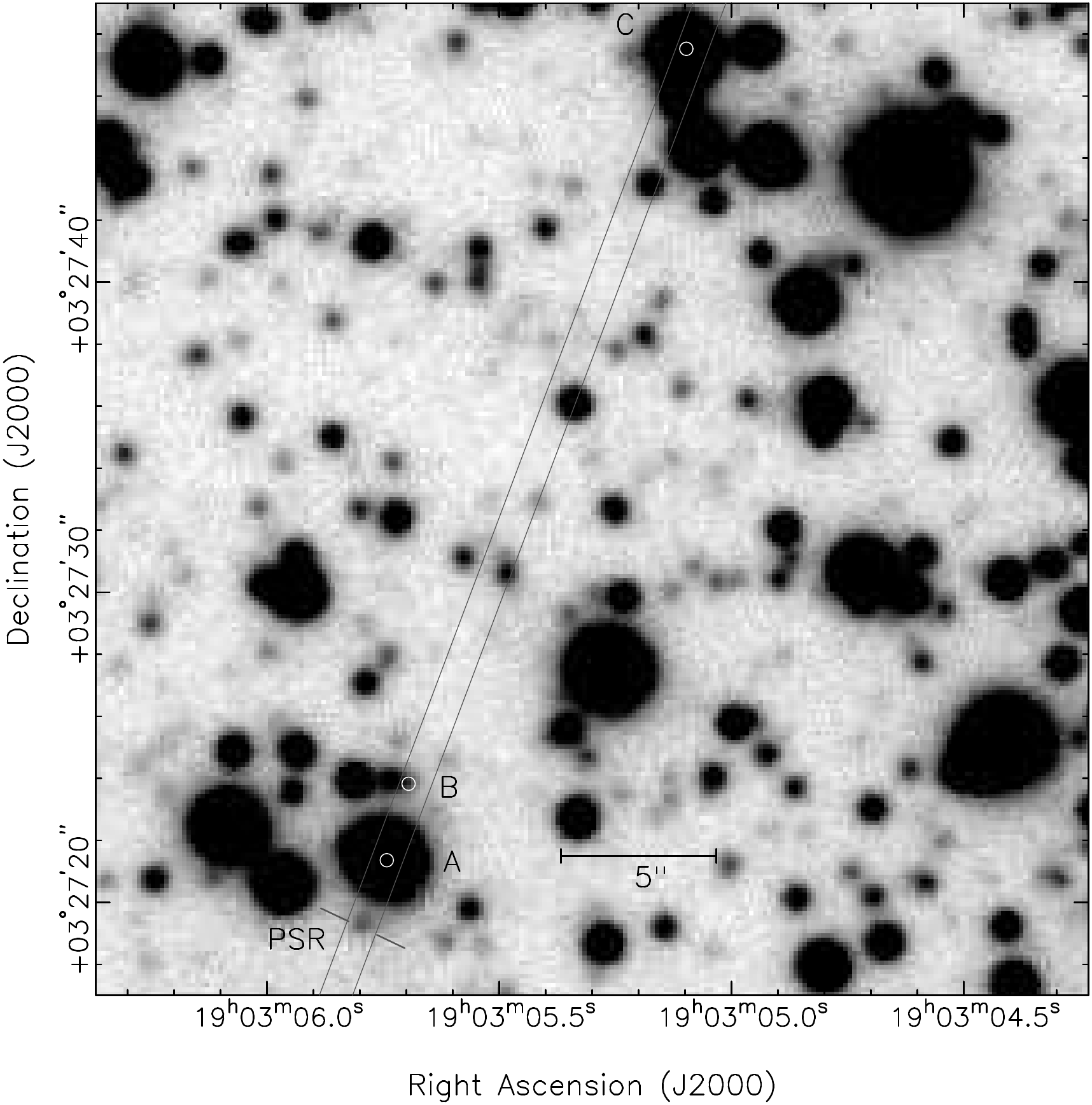} 
  \caption{This $32\arcsec\times32\arcsec$ subsection of a 5\,min SDSS
    $i$-band image taken with GMOS at Gemini North on Sept.\,20th,
    2007 shows the location of the $1\arcsec$ slit. Besides the
    companion to \psr\, denoted with 'PSR', spectra were also
    extracted of stars A, B and C.}
\end{figure}

\begin{figure*}
  \includegraphics[width=17cm]{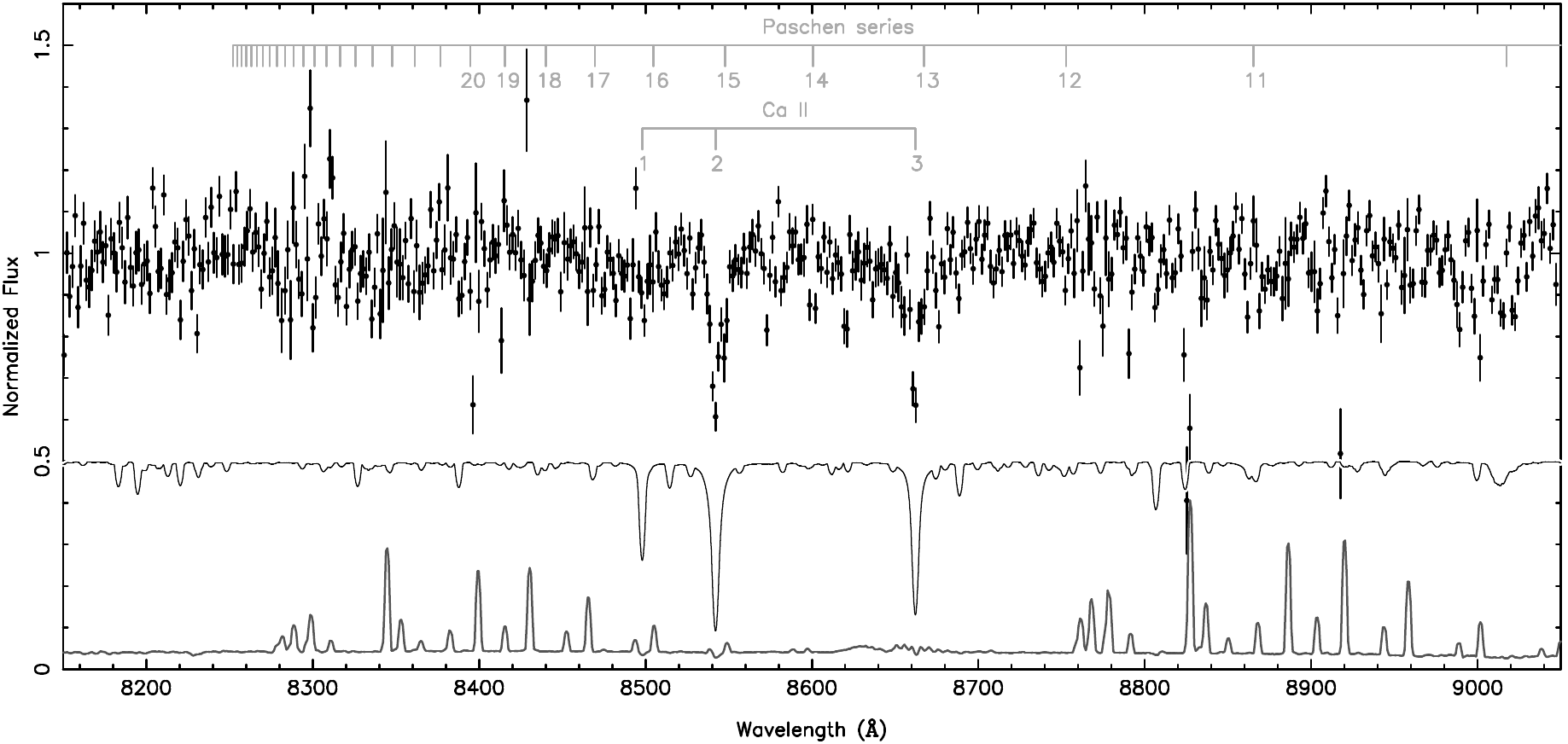}
  \caption{The spectrum of the counterpart of \psr. Shown with error
    bars are the average data from the 8 individual spectra,
    normalised, shifted to zero velocity and binned by a factor of
    2. The best fit normalised spectrum from the library of Munari et
    al.\,(2005) is shown as the solid black line, shifted downwards by
    0.5 units. The average night sky emission spectrum is shown by the
    solid grey line, showing the location of emission lines as regions
    of reduced sensitivity. The vertical scale is arbitrary. The
    location of stellar absorption lines belonging to the
    Ca\,\textsc{ii} triplet and the Paschen series of Hydrogen are
    denoted in grey above the spectrum.}
  \label{fig:spec}
\end{figure*}

Long-slit spectroscopy of the suspected counterpart to \psr\, was
obtained with FORS2 \cite{aff+98}, the low dispersion spectrograph of
ESO's Very Large Telescope. Four spectra were obtained on 2008 June 21,
three on 2008 August 23 and one a day later, on August 24.
All spectra had exposure times of 46\,minutes, and used a $1\asec$
slit combined with the 1028Z holographic grism, providing wavelength
coverage over 7830\,\AA\ to 9570\,\AA. The detectors were read out
with $2\times2$ binning, yielding a resolution of 3.4\,\AA, sampled at
$0.86$\,\AA\,pix$^{-1}$. The slit was placed such that both the pulsar
companion and a bright nearby star were centred on the slit. The
observations were taken during clear and photometric nights, with the
seeing between $0\farcs48$ and $0\farcs72$. The spectral observations
were corrected for bias and flat-fielded using lamp flats.

Spectral extraction is complicated by the bright star, henceforth star
A, located $2\farcs3$ from the pulsar counterpart (see
  Fig.~\ref{fig:finder}). The star is brighter by about 5.6\,mag in
the $I$-band and as a result about 20\% of the detected counts at
the spatial position of the pulsar counterpart are from the wings of
the brighter star. The regular optimal extraction algorithm as
described by Horne~(1986)\nocite{hor86} is not suited to extract
blended spectra as it makes no assumptions about the spatial
profile. Instead, we use a variation on the algorithm by
Hynes~(2002)\nocite{hyn02}, which is based on the optimal extraction
method of Horne~(1986)\nocite{hor86}.

Hynes~(2002)\nocite{hyn02} uses an analytic function to describe the
spatial profile as a function of wavelength. This profile is fitted to
the spectrum of an isolated template source and then used to
simultaneously extract the spectra of the blended sources. We use a
variation on this method. First, instead of using a Voigt function to
describe the spatial profile we use a Moffat~(1969)\nocite{mof69}
function, essentially a modified Lorentzian with a variable
exponent. Secondly, instead of removing the sky contribution before
extraction, we include it in the fit, representing the sky as a first
order polynomial added to the Moffat profiles for each object in the
blend. Finally, the absence of an isolated template source forced us
to use star A as a template. We used an iterative scheme to converge
the properties of the profile as a function of wavelength from star A
while removing the contribution of the pulsar counterpart and another
faint star (star B) that was also part of the blend. In addition to
these three objects, a fourth star (star C) is located 28\asec\,
North-West of \psr\, on the slit. This star was not blended and
extracted normally.

Arc lamp exposures obtained during daytime with the telescope pointing
towards the zenith were used for wavelength calibration. However,
comparison of wavelengths of the night sky emission lines from the
science exposures showed wavelength offsets of up to 0.8\,\AA\ between
different exposures, most likely due to flexure caused by the telescope
pointing away from zenith. To correct for these offsets the wavelength
calibration of June 21st was used to create a secondary line list of
some 60 night sky emission lines in the first spectrum, which was
subsequently used to calibrate the remaining 7 spectra. Typical rms
residuals of these fits were less than 0.05\,\AA.

Figure~\ref{fig:spec} shows the averaged spectrum of the
  counterpart to \psr. The spectrum has a signal-to-noise ratio (S/N)
of 15 to 20 in the wavelength range of 8000\,\AA\ to 9000\,\AA.  The
spectrum is generally featureless except for clear absorption lines
around 8540\,\AA\ and 8660\,\AA. At these wavelengths the most common
spectral lines are those belonging to the Ca\,\textsc{ii} triplet, and
the Paschen series of Hydrogen \citep[see][]{ccg+01}. Though the
Paschen series has lines on these wavelengths, the absence of the
other lines in the series argues for them being due to the Calcium
triplet.

\subsection{Radio timing observations}
\label{sec:tobs}

\begin{figure}
  \centering
  \includegraphics[width=8.6cm]{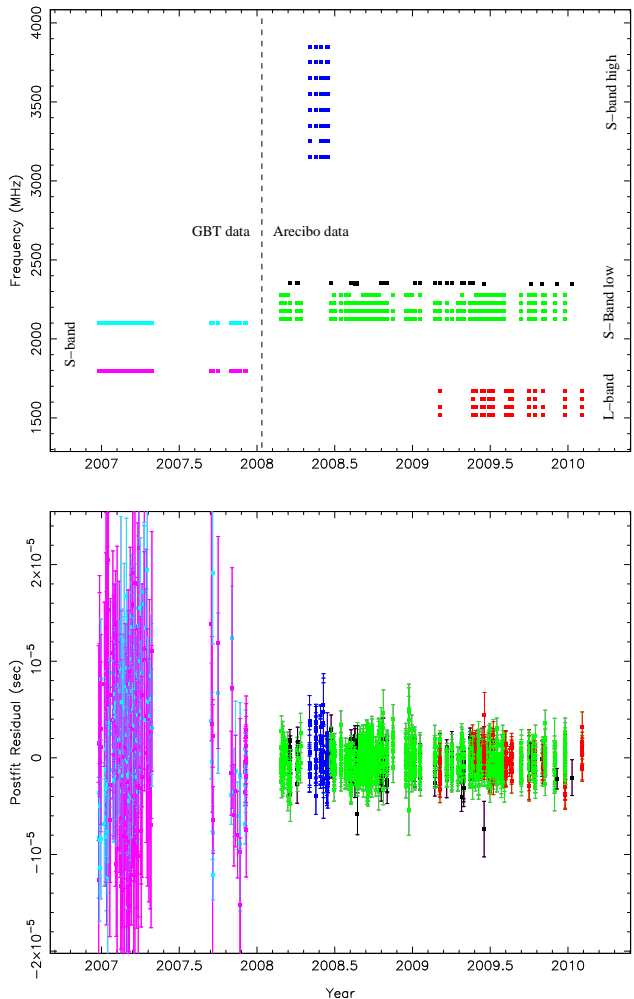}
  \caption{{\em Top:} TOA frequencies versus time (in years). {\em
      Bottom}: Post-fit TOA residuals versus time (in years). The
    different colours indicate different observing systems.  {\em Light
      Blue:} Green Bank SPIGOT data, centred at 2100 MHz, {\em Pink:}
    Green Bank SPIGOT data, centred at 1800 MHz. This was not
    previously taken into account.  {\em Dark blue:} Arecibo WAPP data
    taken with the ``S-high'' receiver.  {\em Light Green:} Arecibo WAPP
    data taken with the ``S-wide'' receiver.  {\em Red:} Arecibo WAPP
    data taken with the ``L-wide'' receiver.  {\em Black:} Arecibo ASP
    data (also taken with the ``S-wide'' receiver).}
  \label{fig:frequencies+residuals}
\end{figure}

We made pulse time of arrival (TOA) measurements of \psr\, using the 305-m
Arecibo radio telescope and the 105-m Green Bank Telescope
from 2006 December through 2010 January; the observation dates and radio
frequencies are summarised in Fig.~\ref{fig:frequencies+residuals}.
The TOAs obtained before 2008 January were described and used by
Champion et al. (2008). We now describe the observational setups used at
the two observatories.

In this re-analysis we use all the TOAs
obtained from processing of data taken with the
Green Bank S-band receiver (with frequency coverage from 1650 to 2250
MHz); these cover most of the year 2007. The data were acquired
using the ``Spigot'' pulsar backend --- a 3-level
autocorrelation spectrometer which samples observing bands of up to
800~MHz \cite{kel+05}. Autocorrelation functions (ACFs) of length 2048
lags were accumulated with 3-level precision and written to disk every
81.92$\mu$s. The ACFs were subsequently Fourier transformed to synthesise
2048-channel power spectra.  We separately analyzed the lower and
upper half of the band, dedispersing each and deriving TOAs using the
methods described in the Supplement of Champion et al. (2008)\nocite{crl+08}.
The TOA uncertainties are 6 $\mu$s in the upper half and 10 $\mu$s in the
lower half of the band. In Champion et al. (2008), only data from the upper
half of the band were used.

The Arecibo data were obtained with four Wideband Arecibo Pulsar
Processors (WAPPs; Dowd, Sisk \& Hagen, 2000)\nocite{dsh00}.
With the ``L-wide'' receiver, the
WAPPs had bandwidth 50 MHz and were centred at 1520, 1570, 1620 and
1670~MHz; each produced 512-lag ACFs sampled with 3-level precision,
accumulated and integrated every 64$\mu$s.  With the ``S-wide''
receiver, the WAPPs had bandwidth 50 MHz and were centred at 2125,
2175, 2225, and 2275 MHz, though the latter was sometimes
disconnected in order to use the Arecibo Signal Processor (ASP,
described below) in its place.  At these frequencies, the WAPPs
produced 256-lag ACFs which were integrated for 32$\mu$s. With the
'S-high' receiver, the WAPPs were configured in dual-board mode,
allowing 100 MHz of bandwidth in each of 8 correlator boards, centred at
100 MHz intervals from 3150 to 3850 MHz.  Each board produced 128-lag
ACFs integrated for 32$\mu$s.

As for the Spigot data, the ACFs from the WAPPs are Fourier
transformed to produce power spectra. These data were then
dedispersed and folded modulo the pulsar's period using a routine
written by one of us (IHS) specifically for this purpose;
this folds the spectra using polynomial coefficients calculated
specifically for the specified observing frequencies. The folding
results in 128-bin pulse profiles every 500 seconds for each WAPP.

During some of the S-wide observations we collected data with the
Arecibo Signal Processor (ASP, Demorest 2007\nocite{dem07}) in
parallel with three WAPPs. The ASP band had a centre frequency of
2350~MHz.  The ASP coherently dedisperses a maximum of 16 4-MHz bands,
for a maximum bandwidth of 64~MHz. These data were then immediately
folded at the pulsar's rotational period, and 512-bin pulse profiles
were stored for each of these bands every 120 seconds. However, because of
the large dispersion measure (DM) of \psr\, we can only coherently
dedisperse 12 of these 4-MHz bands at any given time.  Finally, we add
all the bands and four 120-s integrations to produce pulse profiles with
good S/N.

All pulse profiles are cross-correlated in the Fourier
domain \cite{tay92} with a low-noise template derived from the sum
of the pulse profiles obtained with the same spectrometer and at the same
frequency. From this, we derive a total of 1872 topocentric TOAs.
Both the high-resolution WAPP and ASP data provide
TOA measurements with an average uncertainty of $\sim 1 \mu$s.

In the next step we use the {\sc tempo2} software package
(Hobbs et al. 2006)\nocite{hem06} for TOA analysis.
This applies the clock corrections
intrinsic to the observatory, the Earth rotation data and the
observatory coordinates to convert the TOAs to Terrestrial Time (TT),
as maintained by the Bureau International des Poids et Mesures.
For conversion of TT TOAs to Coordinated Barycentric Time
(TCB) TOAs we used the DE/LE 421 solar system ephemeris
(Folkner, Williams \& Boggs, 2008)\nocite{fwb08}.
The program then minimises the squares of TOA residuals --- 
the difference between the observed and predicted TOAs. We used
the ``DD'' orbital model \cite{dd85,dd86}; this is optimal for
describing eccentric orbits in a theory-independent manner.
The best-fit parameters are listed
in Table 1, their uncertainties are the 1-$\sigma$ values estimated
by {\sc tempo2} (throughout this paper, the quoted uncertainties are
1-$\sigma$, except when stated otherwise). The TOA residuals are displayed in
the bottom plot of Fig.~\ref{fig:frequencies+residuals}. 

We have performed a parallel analysis using the {\sc tempo} software
package\footnote{http://sourceforge.net/projects/tempo/}.
The resulting timing parameters and their uncertainties (as estimated
by {\sc tempo}) are also presented in Table 1. Several
timing parameters are different from those estimated by {\sc tempo2}.
All time-like quantities (spin frequency, orbital period,
projected semi-major axis, time of passage through periastron and
$\omega$, which is strongly covariant with the latter) are different because
{\sc tempo} converts the TOAs to Barycentric Dynamic Time (TDB), a time scale
that is different from TCB. The right ascension and declination are
different because the DE/LE 405 solar system
ephemeris \cite{sta98b} uses an earlier celestial reference frame.
Furthermore, we included in {\sc tempo} a modified version of the
DD orbital model that uses the orthometric parameterization of
the Shapiro delay, as described in \cite{fw10}.
The ``orthometric amplitude'' ($h_3$) and
the ``orthometric ratio'' ($\varsigma$) provide an improved
description of the areas of the ($m_c, \sin i$) plane where the
system is likely to be compared to the ``range'' ($r$) and ``shape'' ($s$)
parameters used in the normal parameterization. In cases where
we can measure other post-Keplerian (PK) parameters, the
new parameterization yields an improved test of general relativity.

Using either {\sc tempo} or {\sc tempo2} there are no apparent trends in the
residuals and the normalised $\chi^2$ is 1.3, which is similar to
what we observe in other MSPs timed with these
observing systems. This means that the timing model provides a complete
description of the observed TOAs, with no significant, unmodelled
effects present. This also means that the 1-$\sigma$ uncertainties
reported by {\sc tempo} and {\sc tempo2} are reasonably accurate
estimates of the real uncertainties.
The non-unity value of the normalised $\chi^2$ likely reflects a slight
underestimation of the TOA uncertainties by the cross-correlation analysis,
but it could also be due to red noise in the pulsar rotation, as suggested
by the marginal (almost 3-$\sigma$) detection of $\ddot{\nu}$.
To account for this,
the timing parameters in Table 1 were obtained after the multiplication
of the TOA uncertainty estimates by a scale factor; this is calculated
separately for each dataset so that its normalised $\chi^2$ is 1.
This scale factor is 1.1 for all the Arecibo data and 1.3 for the GBT data.

\begin{table*}
\begin{center}
\begin{tabular}{l l l}
  \hline
  \multicolumn{3}{c}{{\bf Timing parameters}} \\
  \hline
  Timing Program & {\sc tempo} & {\sc tempo2} \\
  \hline
  Solar System Ephemeris                   \dotfill & DE 405/LE 405 & DE 421/LE 421 \\
  Reference Time Scale                   \dotfill & TDB & TCB \\
  Orbital Model                   \dotfill & Modified DD & DD \\
  Reference Time (MJD)                    \dotfill & 55000 & 55000 \\
  Right Ascension, $\alpha$ (J2000)           \dotfill & 19$^{\rm h}\;$03$^{\rm m}\;$05$\fs$793296(2) & 19$^{\rm h}\;$03$^{\rm m}\;$05$\fs$793213(10) \\
  Declination, $\delta$ (J2000)               \dotfill & 03$\degrees\;$27$\amin\;$19$\farcs$21053(6) & 03$\degrees\;$27$\amin\;$19$\farcs$20911(6) \\
  Proper Motion in $\alpha$, $\mu_{\alpha}$ (mas yr$^{-1}$) \dotfill & $-$2.01(7)   & $-$2.06(7) \\
  Proper Motion in $\delta$, $\mu_{\delta}$ (mas yr$^{-1}$) \dotfill & $-$5.20(12)  & $-$5.21(12) \\
  Spin Frequency, $\nu$ (Hz)                  \dotfill & 465.135245551237(10)  & 465.135238339217(9) \\
  First Derivative of $\nu$, $\dot{\nu}$ ($10^{-15}$ Hz s$^{-1}$) \dotfill & $-$4.0719(2) & $-$4.0719(2) \\
  Dispersion Measure, DM (cm$^{-3}$ pc)       \dotfill & 297.5244(6) & 297.5245(6) \\
  First Derivative of DM (cm$^{-3}$ pc yr$^{-1}$) \dotfill & $-$0.0083(6) & $-$0.0084(6) \\
  Orbital Period $P_b$ (days)                 \dotfill & 95.174117277(14) & 95.174118753(14) \\
  Projected Semi-Major Axis, $x$ (lt-s)       \dotfill & 105.5934628(5) & 105.5934643(5) \\
  Eccentricity, $e$                           \dotfill & 0.436678410(3) & 0.436678409(3) \\
  Longitude of Periastron, $\omega$ ($^\circ$) \dotfill & 141.6531042(2) & 141.6524786(6) \\
  Time of Passage through Periastron, $T_0$ (MJD) \dotfill & 55015.58140451(4) & 55015.58158859(4) \\
  Derivative of $x$, $\dot{x}_o$ ($10^{-15}$lt-s $\rm s^{-1}$) \dotfill & +20(3) & +21(3)\\
  Apsidal Motion, $\dot{\omega}_o$ ($^\circ \rm yr^{-1}$) \dotfill & 0.0002400(2)  & 0.0002400(2) \\
  ``Range'' Parameter of the Shapiro Delay, $r / T_{\odot}$ ($M_{\odot}$) \dotfill &  - & 1.03(3) \\
  ``Shape'' Parameter of the Shapiro Delay, $s$ \dotfill & - & 0.9760(15) \\
  Orthometric Amplitude of the Shapiro Delay, $h_3$ ($\mu$s) \dotfill & 2.602(25) & - \\
  Orthometric Ratio of the Shapiro Delay, $\varsigma$   \dotfill & 0.803(6) & - \\
 \hline
  \multicolumn{3}{c}{{\bf Limits} (not fitted with other timing parameters)} \\
  \hline
  Second Derivative of $\nu$, $\ddot{\nu}$ ($10^{-26}$ Hz s$^{-2}$) \dotfill & 4.8(1.7) & 6.1(1.6) \\
  First Derivative of $e$, $\dot{e}$ ($10^{-16}$ s$^{-1}$)       \dotfill  & 1.4(6) & - \\
  First Derivative of $P_b$, $\dot{P}_b$ ($10^{-12} \rm s s^{-1}$) \dotfill & $-$53(33)  & $-$64(31) \\
  \hline
  \multicolumn{3}{c}{{\bf Derived Parameters}} \\
  \hline
  Spin Period, $P$ (ms) \dotfill & \multicolumn{2}{l}{2.14991236434921(7)} \\
  First Derivative of Spin Period, $\dot{P}$ ($10^{-20} \rm s s^{-1}$) \dotfill & \multicolumn{2}{l}{ 1.88203(11) } \\
  Characteristic Age, $\tau_c$ ($10^{9}$ yr)  \dotfill & \multicolumn{2}{l}{ 1.9} \\
  Dipolar Magnetic Flux Density at the Poles, $B_0$ ($10^8$G) \dots & \multicolumn{2}{l}{ 2.0 } \\
  Galactic Longitude, $l$ \dotfill & \multicolumn{2}{l}{ 37\fdg3363 } \\
  Galactic Latitude, $b$ \dotfill & \multicolumn{2}{l}{ $-$1\fdg0136 } \\
  Distance, $D$ (kpc) \dotfill & \multicolumn{2}{l}{ 6.4$\,\pm\,$1.0} \\
  Total Proper Motion, $\mu$ (mas yr$^{-1}$)  \dotfill & \multicolumn{2}{l}{ 5.60(11) } \\
  Galactic Position Angle of Proper Motion, $\Theta_{\mu}$  \dotfill & \multicolumn{2}{l}{ $ 263.9 \pm 0.8^\circ$ } \\
  \multicolumn{3}{l}{{\bf Velocities}, Solar System Barycentre Reference Frame} \\
  \hspace*{08pt}Transverse velocity, $V_T$ (km s$^{-1}$)  \dotfill & \multicolumn{2}{l}{ 168$\,\pm\,$24 } \\
  \hspace*{08pt}Radial velocity, $\gamma$ (km s$^{-1}$)  \dotfill & \multicolumn{2}{l}{ 42.0$\,\pm\,$4.4 } \\
  \hspace*{12pt}Total 3-D velocity, $V$ (km s$^{-1}$)  \dotfill & \multicolumn{2}{l}{ 174$\,\pm\,$25 } \\
  \multicolumn{3}{l}{{\bf Velocities}, Galactocentric Reference Frame (cylindrical)} \\
  \hspace*{08pt}$V_{\rho}$ (km s$^{-1}$)  \dotfill & \multicolumn{2}{l}{ 17$\,\pm\,$13 } \\
  \hspace*{08pt}$V_{\phi}$ (km s$^{-1}$)  \dotfill & \multicolumn{2}{l}{ $-189\,\pm\,$5 } \\
  \hspace*{08pt}$V_{z}$ (km s$^{-1}$)  \dotfill & \multicolumn{2}{l}{ $-11\,\pm\,$4 } \\
  \hspace*{12pt}Total 3-D velocity, $V$ (km s$^{-1}$)  \dotfill & \multicolumn{2}{l}{ 190$\,\pm\,$5 } \\
  \multicolumn{3}{l}{{\bf Velocities}, Relative to Pulsar Standard of Rest (cylindrical)} \\
  \hspace*{08pt}$V_{\phi}$ (km s$^{-1}$)  \dotfill & \multicolumn{2}{l}{ 28$\,\pm\,$5 } \\
  \hspace*{12pt}Total 3-D velocity, $V$ (km s$^{-1}$)  \dotfill & \multicolumn{2}{l}{ 37$\,\pm\,$9 } \\
  \multicolumn{3}{l}{{\bf Derived Masses and Orbital Orientation}} \\
  Mass Function, $f$ ($M_{\odot}$)   \dotfill & \multicolumn{2}{l}{ 0.139558441(2) } \\
  Orbital Inclination, $i$ ($^\circ$) \dotfill & \multicolumn{2}{l}{ 77.47(15) or 102.53(15) } \\
  Total Mass of Binary, $M_t$ ($M_{\odot}$)   \dotfill & \multicolumn{2}{l}{ 2.697(29)$^b$ } \\
  Companion Mass, $m_c$ ($M_{\odot}$)   \dotfill & \multicolumn{2}{l}{ 1.029(8)$^b$ } \\
  Pulsar Mass, $m_p$ ($M_{\odot}$)   \dotfill & \multicolumn{2}{l}{ 1.667(21)$^b$ } \\
  \hline
\end{tabular}
\caption{\label{tab:parameters}Timing and derived parameters for \psr\, using 
{\sc tempo} and {\sc tempo2} as described in the text.
All parameters are as measured at the Solar System Barycentre. In parentheses we present
the 1-$\sigma$ uncertainties, except in a few cases ($^b$) where we present 99.7\% confidence limits.}
\end{center}
\end{table*}

\section{Results}
\label{sec:results}

\begin{table*}
  \footnotesize \centering
  \caption[]{The celestial position, $J\!H\!K_\mathrm{s}$ magnitudes
    and averaged radial velocities (relative to the solar system
    barycentre) at both epochs of the four stars located on the
    slit. The position and magnitudes are obtained from the data
    described in Champion et al.\,(2008).}
  \label{tab:radvel} \renewcommand{\footnoterule}{}
  \begin{tabular}{lccccccc}
    \hline
    \hline
    ID & $\alpha_\mathrm{J2000}$ & $\delta_\mathrm{J2000}$ & $J$ & $H$ & $K_\mathrm{s}$ & $V_1$ (km\,s$^{-1}$) & $V_2$ (km\,s$^{-1}$)\\
    \hline
    A & $19^\mathrm{h}03^\mathrm{m}05\fs751$ & $+03\degr27\arcmin21\farcs31$ & $14.85\pm0.09$ & $14.44\pm0.10$ & $14.23\pm0.19$ & $42.0\pm0.2$ & $56.3\pm0.2$ \\
    B & $19^\mathrm{h}03^\mathrm{m}05\fs703$ & $+03\degr27\arcmin23\farcs83$ & $19.49\pm0.09$ & $18.52\pm0.10$ & $18.08\pm0.09$ & $49.6\pm6.1$ & $48.6\pm2.7$ \\
    C & $19^\mathrm{h}03^\mathrm{m}05\fs106$ & $+03\degr27\arcmin47\farcs50$ & $15.07\pm0.09$ & $14.52\pm0.10$ & $14.27\pm0.09$ & $45.9\pm0.2$ & $54.3\pm0.4$ \\
    PSR & $19^\mathrm{h}03^\mathrm{m}05\fs803$ & $+03\degr27\arcmin19\farcs18$ & $19.22\pm0.09$ & $18.41\pm0.10$ & $18.03\pm0.09$ & $92.4\pm8.3$ & $20.2\pm5.2$ \\
    \hline
  \end{tabular}
\end{table*}

\subsection{On the nature of the companion}
\label{sec:oresults}
Radial velocities and spectral parameters were determined by comparing
the observed spectra, both individual and averages at each epoch, with
synthetic spectral templates from the spectral library of Munari et
al.\,(2005)\nocite{mscz05}. These models range in effective
temperature $T_\mathrm{eff}$, surface gravity $g$, rotational
velocity $V_\mathrm{rot}$ and metallicity $\mathrm{[M/H]}$ and are
sampled at a resolution of $\lambda/{\Delta \lambda}=20000$. The
wavelength range between 8400\,\AA\ and 8800\,\AA\ was used. To remove
the effects of the finite resolution of the instrument, the synthetic
spectra were convolved with a Gaussian profile with a width equal to
the seeing, and truncated at the width of the slit. For each object,
an iterative procedure was used to first determine the radial
velocities using a starting template, then use those velocities to
shift all spectra to zero velocity and create an average which was
then used to obtain a better matching template for determining more
accurate radial velocities.

\begin{figure}
  \includegraphics[width=8.2cm]{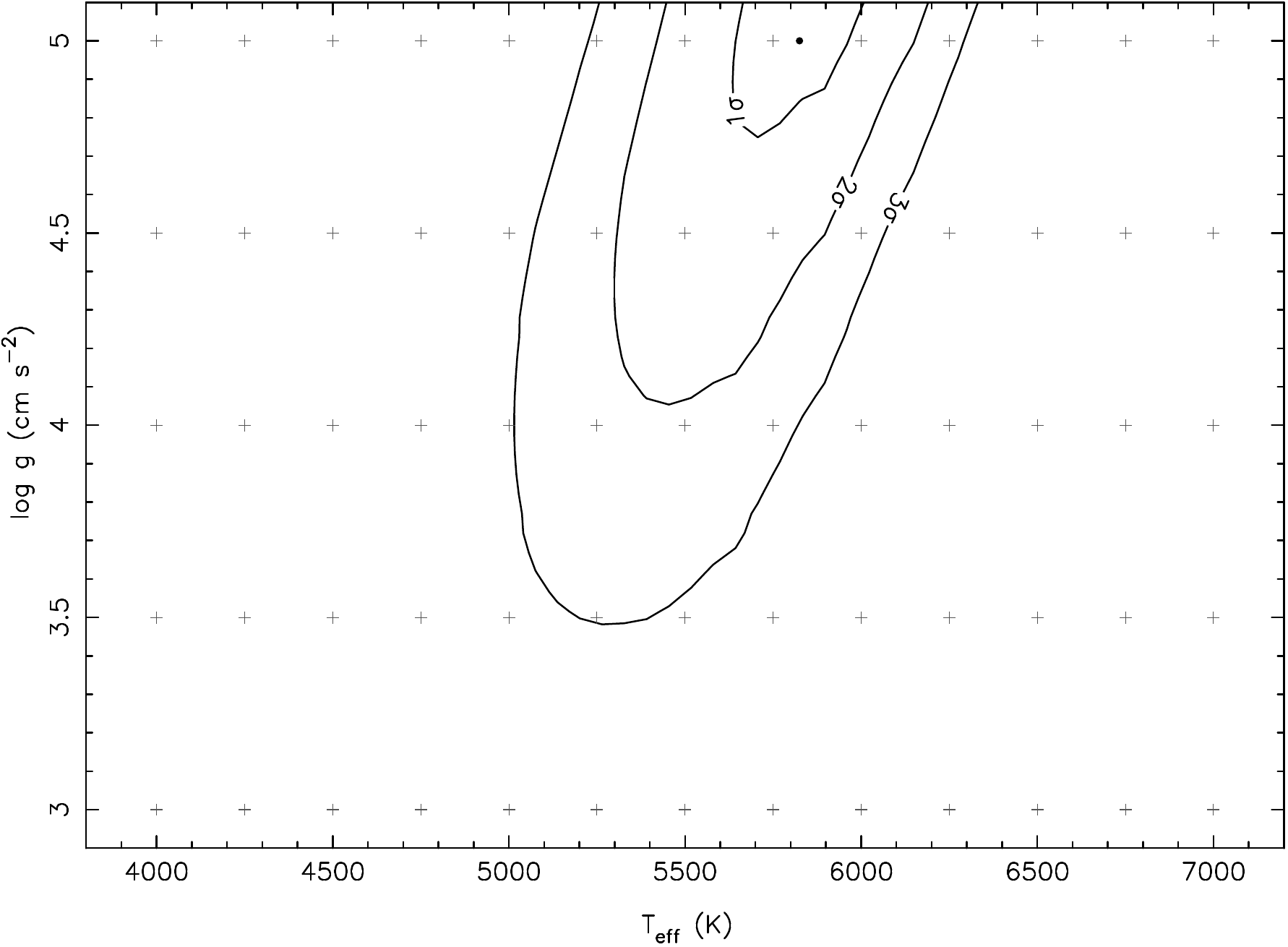}
  \caption{Confidence contours for the effective temperature
    ($T_\mathrm{eff}$) and surface gravity [$\log (g / \rm cm s^{-2})$] for
    the companion of \psr.  }
  \label{fig:tefflogg}
\end{figure}

The comparison of the average spectrum with synthetic spectra from the
library of \cite{mscz05} yields a temperature of
$T_\mathrm{eff}\,=\,5825\pm200$\,K ($1\sigma$) and sets a 2-$\sigma$
lower limit on the surface gravity of $\log g > 4 \rm\,cm\,s^{-2}$,
firmly excluding the possibility of a giant or sub-giant star
(see Fig.~\ref{fig:tefflogg}).  Based
on the shape of the absorption lines we estimate a 3-$\sigma$ upper
limit on the rotational broadening of $v_\mathrm{rot} \sin i_* <
140$\,km\,s$^{-1}$, where $i_*$ is the inclination of the star. The
observed spectrum does not constrain the metallicity. Stellar
evolution models \cite{gbbc00} show that a main sequence (MS)
star with the observed mass and effective temperature of the companion
of \psr\, is consistent with the observed infrared colours for distances
between 6 and 8 kpc, assuming the estimated reddening as a function of
distance from 2MASS stars \cite{crl+08}.  This is in agreement with
the 6.4 kpc distance estimated from the pulsar's DM (see Table 1)
using the \cite{cl02} model of the Galactic electron distribution.
The predicted ages from these models range from 4 to 7\,Gyr.

\subsection{Radial Velocities}

\begin{figure}
  \includegraphics[width=8.2cm]{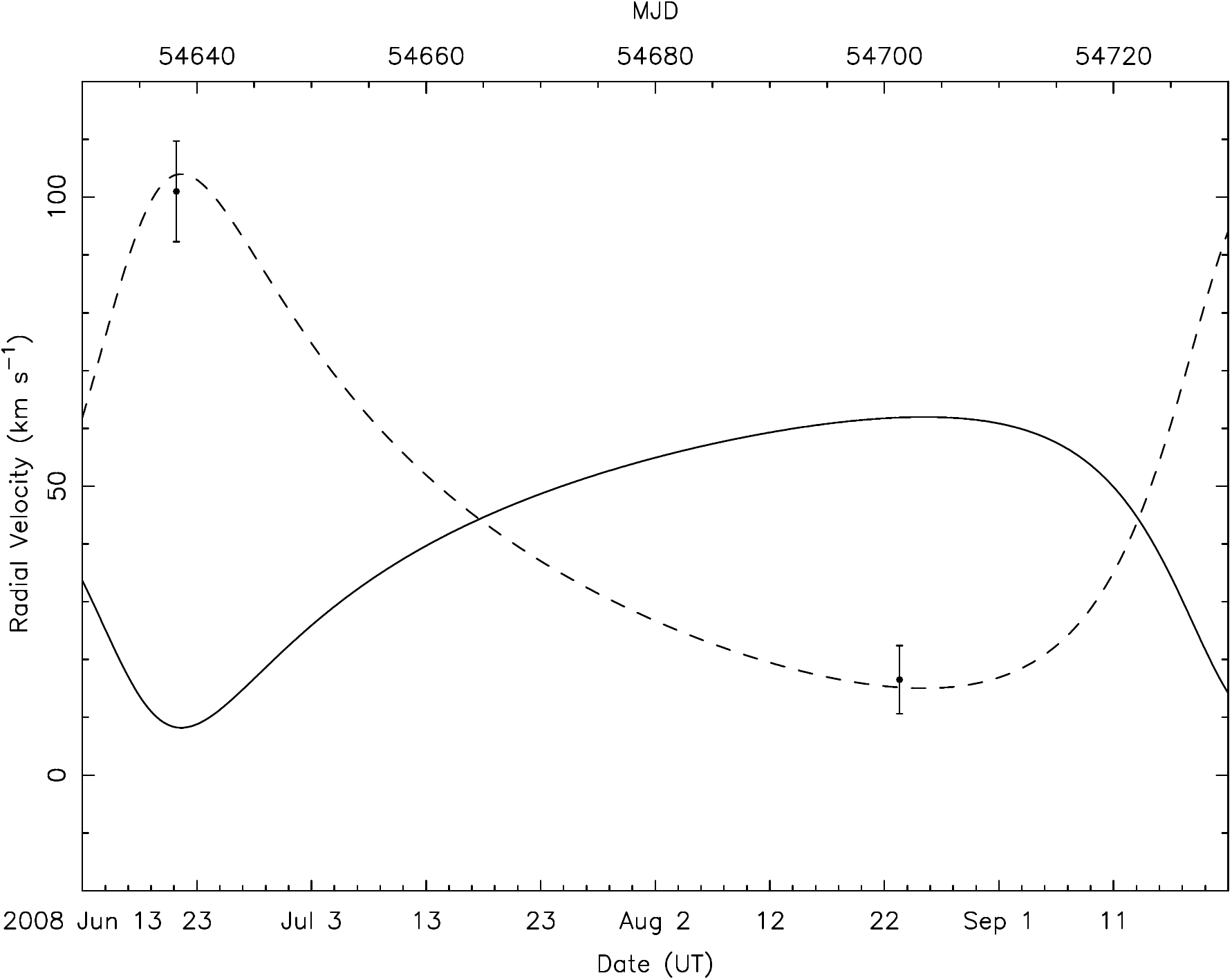}
  \caption{The predicted radial velocities of the pulsar (solid line)
    and companion (dashed line) based on the pulsar timing
    ephemeris. These are vertically offset by the systemic radial
    velocity of the system relative to the solar system barycentre
    $\gamma$, which cannot be determined from the radio
    measurements. The two optical radial velocity measurements are
    shown with their error bars. Their difference is consistent with
    the predictions from the pulsar timing ephemeris, and their
    absolute values allow an estimate of
    $\gamma=44.3\pm4.9$\,km\,s$^{-1}$.  }
  \label{fig:radial}
\end{figure}

 Table~\ref{tab:radvel} shows the radial velocities of the
  pulsar counterpart relative to the Solar System barycentre. They
  are $V_1\,=\,92.4\pm8.3$\,km\,s$^{-1}$ during the first epoch, and
  $V_2\,=\,20.2\pm5.2$\,km\,s$^{-1}$ during the second. The velocity
  difference between the two epochs of
  $V_1-V_2\,=\,72.2\pm9.8$\,km\,s$^{-1}$ deviates from constant
  velocity at the $7.4\sigma$ level. Between the two epochs, the
  averaged radial velocities of stars A and C differ by
  $V_1-V_2=-14.2\pm0.3$\,km\,s$^{-1}$ for star A and
  $-8.4\pm0.4$\,km\,s$^{-1}$ for star C. These differences are caused
  by slight differences in centring of the source on the slit
  \cite{bkkv06}. Based on the location of star A with respect to the
  centre of the slit before and after each exposure, we estimate an
  average offset in radial velocity of about
  $-8.6\pm3.4$\,km\,s$^{-1}$ for the first epoch and
  $3.7\pm2.9$\,km\,s$^{-1}$ for the second epoch. This is comparable
  to the weighted average in $V_1-V_2=-12.1\pm0.3$\,km\,s$^{-1}$ for
  stars A and C.

  Subtracting these radial velocity offsets from the measured
  velocities of the counterpart, we obtain
  $V_1=101.0\pm8.7$\,km\,s$^{-1}$ and $V_2=16.5\pm5.9$\,km\,s$^{-1}$,
  and a velocity difference of $V_1-V_2=85\pm11$\,km\,s$^{-1}$. This
  is consistent with the $V_1-V_2\,=\,88.72$\,km\,s$^{-1}$ difference
  in radial velocity predicted from the pulsar timing ephemeris. This
  confirms that the optical counterpart to \psr\, is the object in the
  95-day orbit around the pulsar. Fig.~\ref{fig:radial} shows the
  radial velocity measurements of the companion and the radial
  velocity predictions based on the orbital parameters determined from
  pulsar timing. This provides an independent estimate of the mass
  ratio ($R\,=\,1.55 \pm 0.20$) and the systemic radial velocity of
  the binary, $\gamma\,=\,44.3\pm4.9$\,km\,s$^{-1}$, which cannot be
  derived from the radio timing.

\subsection{Shapiro delay} \label{sec:shap}

\begin{figure}
  \centering
  \includegraphics[width=8.6cm]{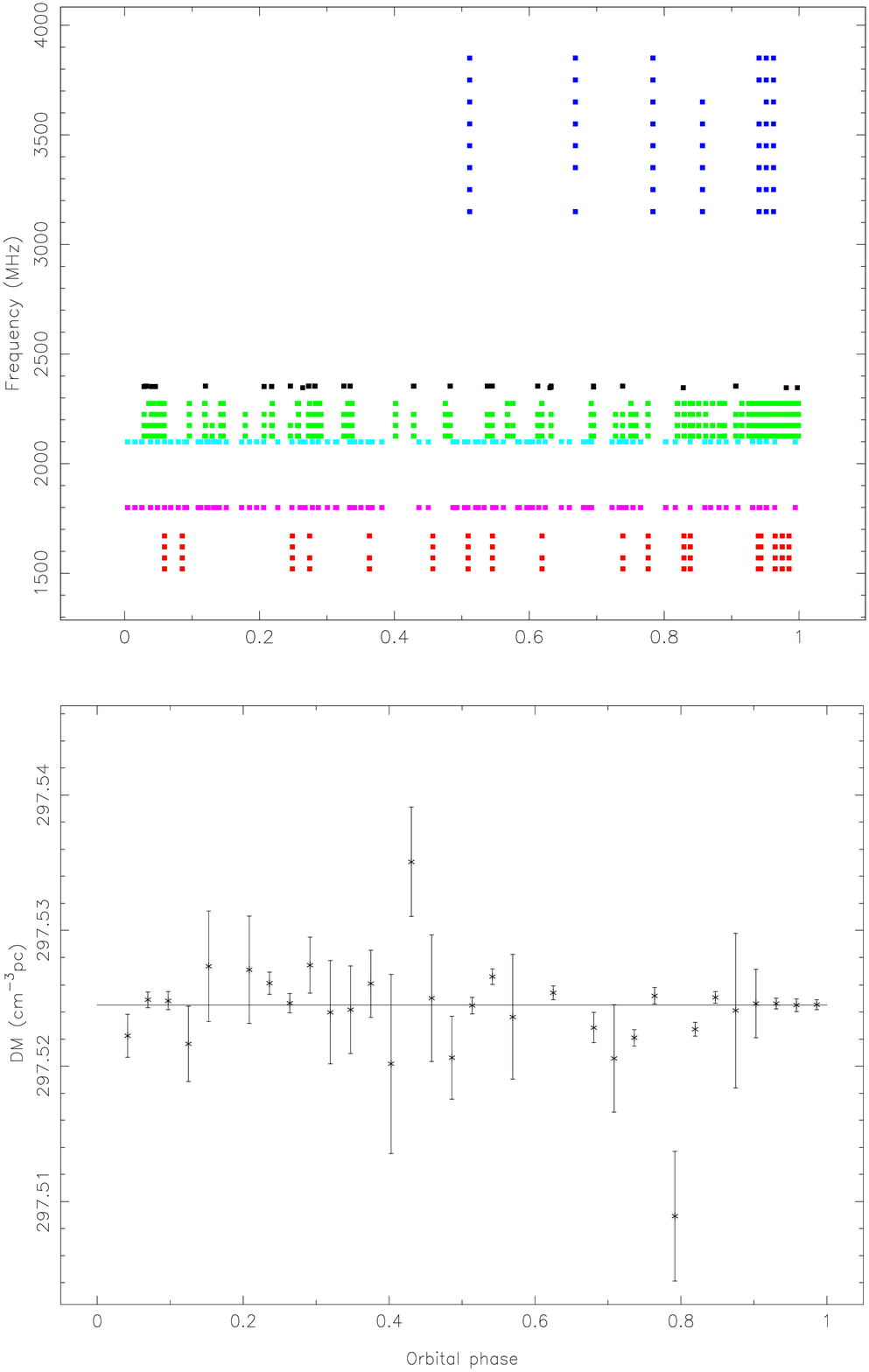}
  \caption{{\em Top}: Radio frequency of TOAs as a function of orbital
    phase.  The orbital phases with the larger number and frequency
    spread of observations are those for which the DM measurements
    should be better.  These are best at an orbital phase of 0.95,
    where superior conjunction happens.  {\em Bottom:} DM
    as a function of orbital phase.  All TOAs were divided in
    $10^\circ$ orbital phase bins and a DM was derived for them,
    keeping all other parameters fixed.}
  \label{fig:DMs}
\end{figure}

Since the companion is a non-degenerate star,
it could in principle have a strong stellar wind, which should
be detectable as a variation of DM as a function of orbital phase. Such
a signal would produce a distortion in our measurement of the
Shapiro delay. In Fig.~\ref{fig:DMs} we display the DM averaged over 36 bins
of the pulsar's mean anomaly, after correction of the long-term DM variations.
For some of these intervals we have smaller amounts of data or a small
range of frequencies. If these result in DM determinations with uncertainties
greater than $ 0.01\,\rm cm^{-3}\,pc$, they are not depicted.
We detect no DM variations greater than $0.001\,\rm cm^{-3}\,pc$.
This means that the mass estimates derived from the Shapiro delay
are accurate, within their uncertainties.

The better timing precision, larger number of TOAs,
longer data span and optimised orbital coverage of our new dataset
not only improve previous measurements of relativistic effects
(namely the observed apsidal motion $\dot{\omega}_o$ and the
$s$ parameter of the Shapiro delay, see Champion et al. 2008)
but also allow a precise measurement of the $r$ parameter of
the Shapiro delay, in addition to a measurement of the proper
motion $\mu$ and the variation it causes on the apparent size of
the orbit $\dot{x}$ (see Table 1). We now discuss
the relevance of these parameters.

\subsubsection{Shapiro delay and component masses}
\begin{figure}
  \centering 
  \includegraphics[width=8.6cm]{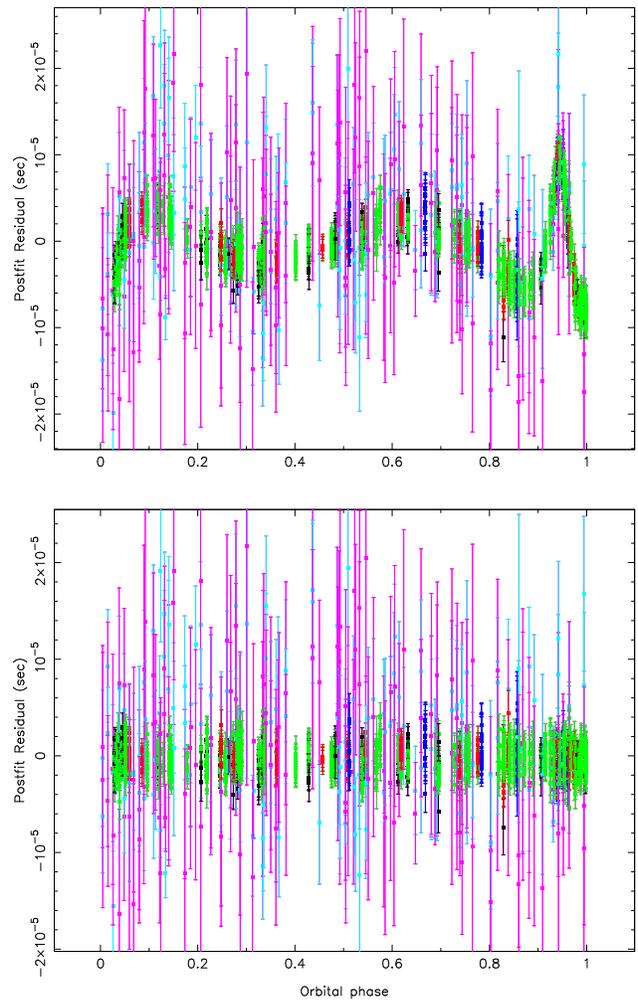}
   \caption{{\em Top:} Post-fit TOAs versus orbital phase. The Shapiro
     delay was not taken into account, but all Keplerian parameters
     were fit. {\em Bottom}: Post-fit TOAs versus orbital phase, with
     Shapiro delay taken into account. The different colours indicate
     different observing systems, as in
     Fig.~\ref{fig:frequencies+residuals}.  }
   \label{fig:residuals_orbit}
\end{figure}

With the previously available timing data it was not possible to
measure the companion mass $m_c$ from the Shapiro delay 
alone. The precise (and frequent) Arecibo timing has completely
changed this situation, providing a very clear Shapiro delay
signal (see Fig.~\ref{fig:residuals_orbit}).
Assuming that general relativity (GR) is the
correct theory of gravity, we obtain
\begin{equation}
\frac{m_c}{M_{\odot}} = \frac{r}{T_{\odot}} \equiv \frac{h_3}{T_{\odot} \varsigma^{3}}  = 1.03 \pm 0.03,
\end{equation}
where $T_{\odot} = G M_{\odot} / c^3 = 4.925490947 \mu$s is the solar
mass $M_{\odot}$ in time units. As usual $G$ is Newton's gravitational
constant and $c$ is the velocity of light.
The orbital inclination derived from $\varsigma$ is $i = (77.4 \pm
0.4)^\circ$ or $(102.6 \pm 0.4)^\circ$.  Given the mass function $f$,
these values imply a pulsar mass $m_p = (1.67 \pm 0.08) M_{\odot}$, a
total binary mass $M_t = (2.70 \pm 0.11) M_{\odot}$ and a mass ratio
$R_s = 1.62 \pm 0.03$.

\subsubsection{Estimating the uncertainties}
The uncertainties quoted above were estimated using the Bayesian
technique described by \cite{sna+02}. We assume that $m_c$ and $\cos
i$ have constant {\em a priori} probability.  For each point in a grid
of $m_c$, $\cos i$ values we calculate (trivially) the $r$ and $s$
parameters to describe the Shapiro delay, assuming that general
relativity is the correct theory of gravitation.

We then fit a timing solution similar to that of Table 1 to the TOAs
but keep ($r$, $s$) fixed. We record the resultant $\chi^2$ and
calculate the probability density for that ($m_c,\cos i$) space using
\begin{equation}
p (m_c, \cos i) =
\exp \left(- \frac{\chi^2(m_c, \cos i) - \chi^2_{\rm min}}{2} \right),
\label{eq:chi2}
\end{equation}
where $\chi^2_{\rm min}$ is the minimum value of $\chi^2$ in the whole
map. The contour levels that include the 68.3, 95.4 and 99.7\%
of all probability are displayed in Fig.~\ref{fig:masses} as the thin
contours. We then project the 2-D probability density function
(PDF) into the $m_c$ and $\cos i$ axes to calculate lateralised 1-D PDFs.

We translate the 2-D PDF of the ($\cos i, m_c$)
space to a 2-D PDF of the ($m_p, m_c$) space using the mass function $f$
\cite{lk05}. This is then lateralised onto the $m_p$ axis, resulting
in a 1-D PDF for the pulsar mass. It is from the 1-D PDFs that we derive
the uncertainties for $\cos i$, $m_p$ and $m_c$.

\subsection{Orientation of the orbit and component masses}
\label{sec:om}
\subsubsection{Kinematic effects: orbital orientation}
The additional, previously undetermined parameters
in Table 1 are the proper motion (with total magnitude $\mu$ and
position angle $\Theta_{\mu}$) and the apparent variation of the size
of the orbit, $\dot{x}_o$; this is given by
\begin{equation}
\frac{\dot{x}_o}{x} = - \mu  \cot i  \sin (\Theta_{\mu} - \Omega) + \frac{\mu^2 D + a_l}{c} + \frac{\dot{x}_s}{x} + \frac{\dot{a}}{a},
\label{eq:dotx}
\end{equation}
where
$i$ is the orbital inclination,
$\Omega$ is the position angle of the line of nodes,
$\dot{x}_s$ is the variation of $x$ due to an intrinsic variation of $i$,
$\dot{a}$ is a variation of the semi-major axis $a$,
$D$ is the distance to the pulsar and
$a_l$ is the difference of Galactic accelerations of the binary and
the solar system, projected along the line of sight.

The first term is by far the largest; it is an effect of the proper motion,
which constantly changes the viewing geometry. If this causes a change
of $i$ as viewed from Earth then there will be
a secular change of the projected size of the orbit, $x = a \sin i /
c$ \cite{ajrt96,kop96}.

The second term describes the changing Doppler shift of the
binary relative to the Solar System barycentre. For a nominal DM distance
of 6.4 kpc, this amounts to $- 4.6 \times 10^{-20} \rm s^{-1}$. This is three
orders of magnitude smaller than the measurement error; this
term is thus ignored in the following discussion. The third term is due to
any real changes in the orbital plane of the binary system.
The only likely contribution to this is from spin-orbit coupling
\cite{sb76,lbk95}; the resulting $\dot{x}_s$ should be one order
of magnitude smaller than the measurement error (see detailed discussion in
\S~\ref{sec:spin-orbit}). Finally, the fourth term is due to a real change
in the size of the orbit. The prediction for $\dot{a}/a$ caused by
gravitational wave emission is of the order of $10^{-26}\rm s^{-1}$;
many orders of magnitude too small.
Therefore, only the first term is likely to give
a significant contribution to the observed $\dot{x}$.

In this case, for each
of the two possible values of $i$ discussed above there are two
possible $(\Theta_{\mu} - \Omega)$ solutions; these are depicted in
Fig.~\ref{fig:orbital_orientation}
 by the intersection of the $\dot{x}$ and $\varsigma$ lines.
However, because of the uncertainty in the measurements of $\dot{x}$
and $\varsigma$, these two pairs of solutions merge into two wide
areas centred at $(\Theta_{\mu} - \Omega) \sim 90^\circ$ for $i \sim
102.6^\circ$ and $(\Theta_{\mu} - \Omega) \sim 270^\circ$ for $i \sim
77.4^\circ$.

\subsubsection{Kinematic effects: apsidal motion and masses}
The observed apsidal motion $\dot{\omega}_{\rm o} = (86.38 \pm 0.08)
\asec/ \rm century$ is, in principle, a combination of the
relativistic apsidal advance $\dot{\omega}_r$; a kinematic
contribution $\dot{\omega}_k$ due to the changing viewing geometry of
the system; and a contribution caused by spin-orbit coupling
$\dot{\omega}_s$:
\begin{equation}
\dot{\omega}_o = \dot{\omega}_k + \dot{\omega}_r + \dot{\omega}_s.
\end{equation}
We will now consider each of these terms in detail.

The $\dot{\omega}_k$ is caused by the change
of viewing geometry due to the proper motion $\mu$ \cite{kop96}:
\begin{equation}
\dot{\omega}_k = \mu'  \cos (\Theta_{\mu} - \Omega),
\label{eq:dotomegak}
\end{equation}
where $\mu' = \mu / \sin i = 0\farcs56 / \rm century$.  Given the
aforementioned
uncertainty of $\Theta_{\mu} - \Omega$ it is impossible at the moment
to know the exact value of $\dot{\omega}_k$, but it has a very sharp
upper limit given by $|\cos (\Theta_{\mu} - \Omega )| \leq 1$.  This
means that uncertainty in $\dot{\omega}_r$ caused by the lack of
precise knowledge of $\dot{\omega}_k$ is at most equal to $\pm \mu' =
0\farcs56 / \rm century$. This is more than six times larger than the
current uncertainty in the measurement of $\dot{\omega}_o$.

This affects the precision of our estimate of the total mass of the
system. If $\dot{\omega}_r = \dot{\omega}_o$,
then the total system mass could be determined directly using
\begin{equation}
\label{eq:omdot_r}
\dot{\omega}_r = 3  n_b^{5/3} \frac{(T_{\odot} M_t)^{2/3}}{1 - e^2},
\end{equation}
where $n_b = 2 \pi / P_b$ is the orbital frequency and $P_b$ is the
orbital period (e.g. Lorimer \& Kramer, 2005)\nocite{lk05};
the result would be $M_t = (2.697 \pm 0.004) M_{\odot}$.
Because of the uncertainty of $\dot{\omega}_k$ and
$\dot{\omega}_r = \dot{\omega}_o - \dot{\omega}_k$, we obtain
instead $M_t = (2.697 \pm 0.029) M_{\odot}$ (99.7\% confidence
limit, see \S~\ref{sec:masses}).

\subsubsection{Further uncertainty estimates}
\label{sec:serious_uncertainty_estimates}
To estimate these uncertainties rigorously,
we have made a second $\chi^2$ map that uses the same principles
described above, but includes a third dimension, the difference in the
position angles of the proper motion and line of nodes $\Theta_{\mu} -
\Omega$. This map makes use of the extra information provided by the
measurements of $\dot{x}$ and $\dot{\omega}$.

As before, from the values of $m_c$ and $\cos i$ at each point we
calculate $r$ and $s$. But now we also use the mass function $f$ to
calculate $m_p$. From the total mass $M_t = m_p + m_c$ we calculate
 $\dot{\omega}_r$ using eq.~\ref{eq:omdot_r}.
From the values of $\cos i$ and $(\Theta_{\mu} - \Omega)$ at each point we
calculate $\dot{x}$ using eq.~\ref{eq:dotx} and $\dot{\omega}_k$ using
eq.~\ref{eq:dotomegak}.  This is made using only the best value for
the proper motion $\mu$, which has a relatively small uncertainty.

\begin{figure*}
  \includegraphics[width=17cm]{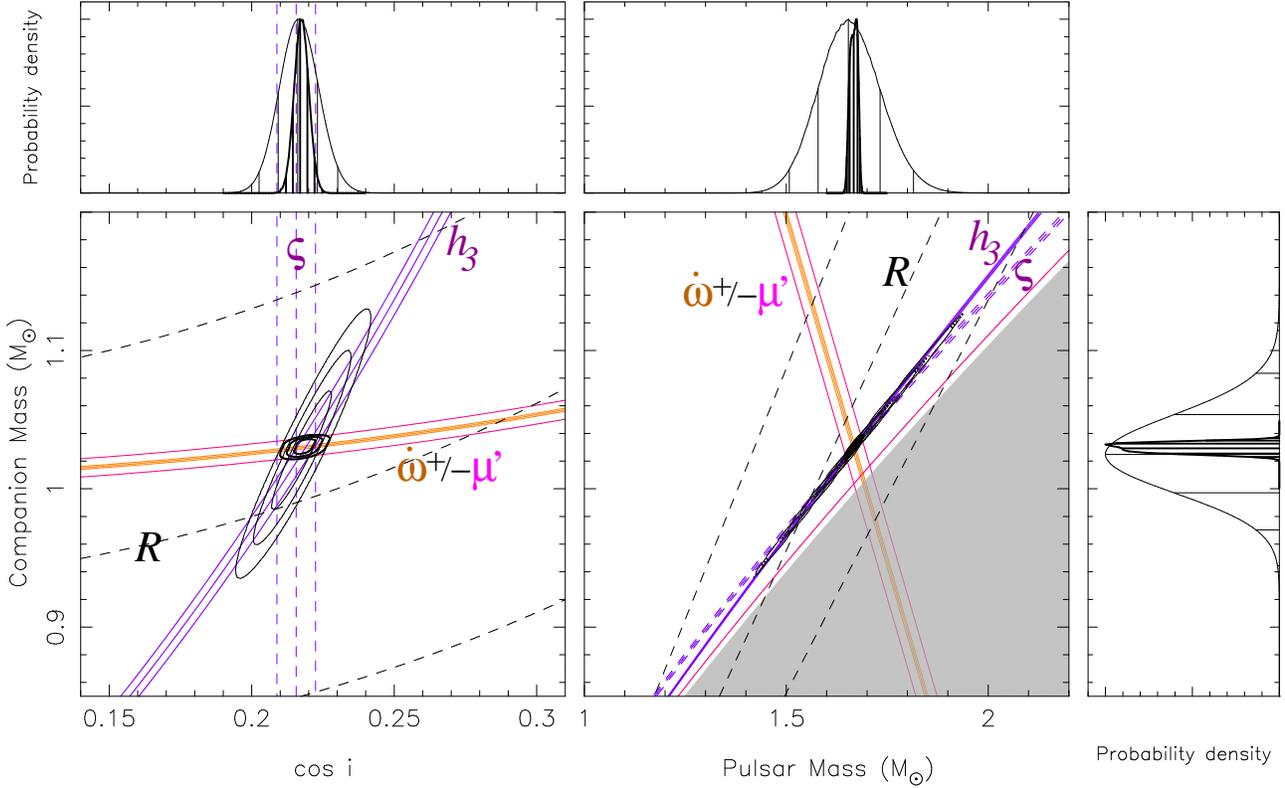}
  \caption{Companion mass as a function of $\cos i$ and $m_p$. The
    thick contour levels are derived from a 3-D $\chi^2$ map of the
    $\cos i$, $m_c$ and $\Omega$ space (where $\Omega$ is the position
    angle of the line of nodes) and then collapsed on the planes
    represented in the figure (see text for details). The thin contour
    levels represent a 2-D $\chi^2$ map of the $\cos i - m_c$ space
    calculated taking only the Shapiro delay into account. The lines
    represent the constraints derived from the spectroscopic mass
    ratio ($R$, black dashed), the apsidal motion ($\dot{\omega}$,
    solid orange - here with increased uncertainty due to the proper
    motion, solid pink), the harmonic amplitude ($h_3$) and harmonic
    ratio ($\varsigma$) of the Shapiro delay (in purple) and finally
    an upper limit on the inclination given by $\dot{x}$ (pink solid
    line). The gray area in the mass-mass diagram is excluded by $\sin
    i \leq 1$. In the marginal plots we can see that the 1-D
    probability distribution functions for the pulsar and companion
    masses are much narrower when the apsidal motion (even with
    uncertainty caused by the proper motion) is taken into account (thick
    lines), but this assumes that there are no significant classical
    contributions to $\dot{\omega}$. The latter must be $< 2.3 \asec/
    \rm century$ (1-$\sigma$) given the agreement between the $h_3$,
    $\varsigma$ and $\dot{\omega} \pm \mu '$ bands.}
  \label{fig:masses} 
\end{figure*}

\begin{figure*}
  \includegraphics[width=17cm]{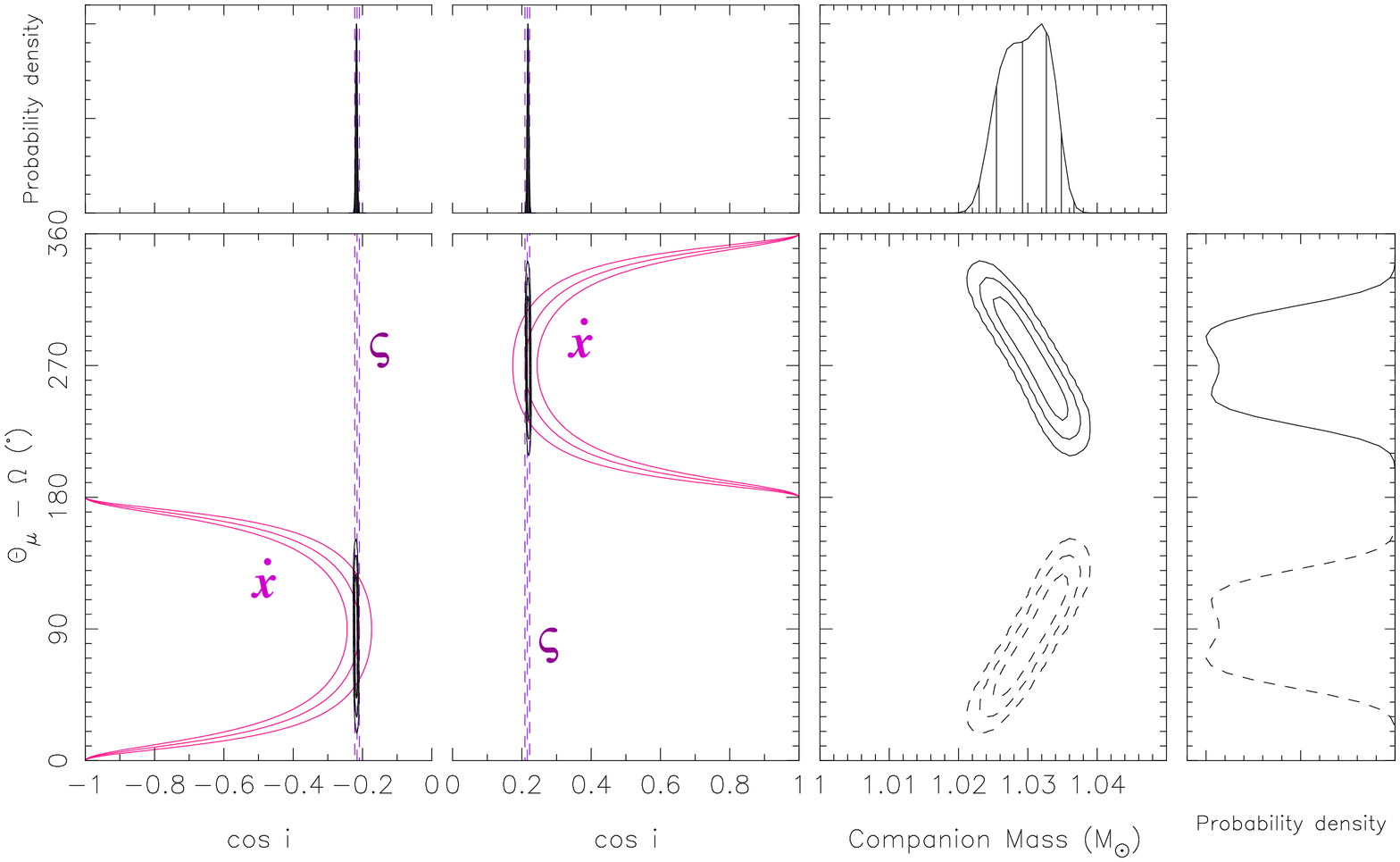}
  \caption{Confidence contours for $\cos i$ and $m_c$ as a function of
    the orientation of the line of nodes relative to the direction of
    the proper motion $\Theta_{\mu} - \Omega$. The contour levels
    include 66.3, 95.4 and 99.7\% of the total probability, which is
    derived from a 3-D $\chi^2$ map of the ($\cos i, m_c, \Theta_{\mu}
    - \Omega$) space and then collapsed onto the planes represented in
    the figure. The dashed contour lines indicate $90^\circ < i <
    180^\circ$. The vertical purple dashed lines are the constraints
    introduced by the harmonic ratio of the Shapiro delay
    ($\varsigma$), the pink solid curves are the constraints derived
    from the measurement of $\dot{x}$. In the marginal plots are
    displayed the 1-D probability distribution functions. We can see
    that there is still a large uncertainty in $(\Theta_{\mu} -
    \Omega)$. This introduces a large uncertainty in the mass
    estimates (of which $m_c$ is shown in the plot).  }
  \label{fig:orbital_orientation}
\end{figure*}

We then fit the resulting timing solution to the TOAs, but keep the
four computed parameters fixed ($r$, $s$, $\dot{x}$ and $\dot{\omega}
= \dot{\omega}_r + \dot{\omega}_k$).  We record the resultant $\chi^2$
and calculate the probability density for that point of the ($m_c$,
$\cos i$, $\Theta_{\mu} - \Omega$) space using an expression similar
to eq.~\ref{eq:chi2}.  This 3-D ``cube'' of probability density is
then projected onto its faces: the ($\cos i,
m_c$) space (see thick contours in Fig.~\ref{fig:masses})
and the ($\cos i, \Theta_{\mu} - \Omega$)
and ($m_c, \Theta_{\mu} - \Omega$) spaces 
(Fig.~\ref{fig:orbital_orientation}). We then project this
3-D PDF onto the three axes, calculating lateralised 1-D PDFs for
$m_c$, $\cos i$ and $\Theta_{\mu} - \Omega$.  It is from these that we
derive the uncertainties of these parameters.

In the  ($m_c, \Theta_{\mu} - \Omega$) panel of 
Fig.~\ref{fig:orbital_orientation} we can see that
if $\Theta_{\mu} - \Omega$ were more precisely known, we would have a
much better estimate of $m_c$. This is the graphical demonstration that
the kinematic effects  are now the main source of uncertainty in the
measurement of $m_c$ and $m_p$, not the uncertainty intrinsic to the
measurement of $\dot{\omega}$. We can also see that as the precision
of $\dot{x}$ improves, we will have a total of four possible values
for $\Theta_{\mu} - \Omega$, two for each possibility of $\cos i$.
The two solutions closer to the centre of Fig.~\ref{fig:orbital_orientation}
($\Theta_{\mu} - \Omega = 180^\circ$) will have a large value of
$m_c$ (and $m_p$), since, according to eq.~\ref{eq:dotomegak}, for this
orbital orientation $\dot{\omega}_k < 0$, therefore
$\dot{\omega}_r > \dot{\omega}_o$. The opposite will
be true for the two solutions closer to $\Theta_{\mu} - \Omega = 0^\circ$.
Unless we can independently determine $\cos i$ or $\Theta_{\mu} - \Omega$,
we will have two degenerate values for $m_c$ and $m_p$. In principle,
this degeneracy can be lifted if we can measure the orbital annual parallax.
This should be possible with a fourfold improvement in timing precision.

\subsubsection{Contribution from spin-orbit coupling}
\label{sec:spin-orbit}
Thus far we have assumed that the spin-orbit contributions to $\dot{\omega}$
and $\dot{x}$ are negligible. Comparing eq. 7.54 and following in
\cite{wil93} with the more general equation in \cite{wex98} we find
for $\dot{\omega}_s$:
\begin{equation}
\dot{\omega}_s = 3 \pi J_2 \frac{R_c^2}{P_b a^2 (1 - e^2)^2}
\left( 1 - \frac{3}{2} \sin^2 \theta + \cot i \sin \theta \cos \theta \cos \Phi_0 \right),
\end{equation}
where $a$ is the semi-major axis of the orbit of the companion as seen
from the pulsar, $R_c$ is the companion radius (which we assume to be
one solar radius), $P_b$ and $e$ are as listed in Table 1, $J_2$ is
the quadrupolar moment of the star, $\theta$ is the angle between the
orbital and companion's angular momentum and $\Phi_0$ is the longitude
of the ascending node in a reference frame defined by the total
angular momentum vector.  For \psr , we obtain a maximum value of
$\dot{\omega}_s = J_2\,\times \, 7.9 \times 10^{5}$ \asec / \rm
century.

As we have shown in \S~\ref{sec:oresults},
the companion star has a mass, age and temperature similar to that of
the Sun. Such stars should rotate slowly like the Sun \cite{iab+09};
the latter has a rotational velocity of $v_\mathrm{rot, \odot} =
2$\,km\,s$^{-1}$.  This should result in a quadrupolar moment similar
to that of the Sun, $J_{2, \odot} \sim 1.7 \times 10^{-7}$. For this
value of $J_2$ we get $\dot{\omega}_s\,\sim \,0\farcs013/ \rm
century$, about seven times smaller than the measurement uncertainty
of $\dot{\omega}_o$.

However, because the companion's rotation is not well constrained by
the optical measurements (which indicate $v_\mathrm{rot}\sin i_* <
140$\,km\,s$^{-1}$, 3-$\sigma$, see \S~\ref{sec:optical}), we can
use the agreement between $h_3$, $\varsigma$ and $\dot{\omega}_r$ to
constrain $\dot{\omega}_s$ assuming that GR is the
correct theory of gravity. The minimum total mass compatible (at the
1-$\sigma$ level) with $h_3$ and $\varsigma$ is $2.59 M_{\odot}$. The
corresponding minimum $\dot{\omega}_r$ ($84.1 \asec/ \rm century $)
can then be used to derive a 1-$\sigma$ limit of $\dot{\omega}_s =
\dot{\omega}_o - \dot{\omega}_{r, \rm min} \,<\,2.3 \asec/ \rm
century$ (here we are assuming a median expected value of $0$ for
$\dot{\omega}_k$).  Future optical/near infrared measurements might be
able to better constrain $v_\mathrm{rot}$. If this is small then the
agreement of the 3 different PK parameters can be used directly
as a test of general relativity.

To calculate $\dot{x}_s$, we use eq. 81 of \cite{wex98} to derive
\begin{equation}
\frac{\dot{x}_s}{x} = 3 \pi J_2 \frac{R_c^2}{P_b a^2 (1 - e^2)^2} \cot i \cos \theta
\sin \theta \sin \Phi_0.
\end{equation}
Using the parameters above, we obtain
$|\dot{x}_s/x| < 2.2 \times 10^{-18}\rm\,s^{-1}$,
which is one order of magnitude below the measurement error.
Therefore, an improved measurement of $\dot{x}$ will likely improve the
determination of the orbital orientation.

\subsubsection{Component masses and orbital inclination}
\label{sec:masses}
From the 3-D map, we obtain the following parameters:
$i\,=\,(77.47 \pm 0.15)^\circ$ or $(102.53 \pm 0.15)^\circ$ (1-$\sigma$),
$M_t\,=\,(2.697 \pm 0.029)\,M_{\odot}$,
$m_c\,=\,(1.029 \pm 0.008)\,M_{\odot}$,
$m_p\,=\,(1.667 \pm 0.021)\,M_{\odot}$
and $R\,=\,1.620 \pm 0.008$.
The underlying probability distribution functions for these parameters
have a shape that is very different from a Gaussian curve
(see Fig.~\ref{fig:masses}), therefore it is not very meaningful to
refer to $\sigma$ limits. For all the previous values we indicated
instead 99.7\% confidence limits. 
These values and their uncertainties can be understood qualitatively as
the result of the intersection of the 1-$\sigma$ bands of
$h_3$ and $\dot{\omega}_r$. Their intersection results in
measurement of $i$ that is more precise than the value
derived from $s$ (or $\varsigma$). Such a result cannot be understood using
the regular $r,s$ parameterization of the Shapiro delay.

These values are entirely consistent with those
derived solely from the Shapiro delay in \S~\ref{sec:shap},
but much more precise, despite the uncertainty introduced by
$\dot{\omega}_k$.

\section{Implications}
\label{sec:discussion}
\subsection{Neutron star mass}
\label{sec:massimplications}
The recent mass measurement of PSR~J1614$-$2230
($m_p = (1.97 \pm 0.4) M_{\odot}$, 1-$\sigma$, see Demorest et al.
2010)\nocite{dpr+10} rules out the presence of hyperons, bosons and
free quarks at densities comparable to the nuclear saturation density,
and demonstrates that NSs can be stable at masses well above
the Chandrasekhar mass. The mass measurement of \psr\,
($m_p = (1.667 \pm 0.021) M_{\odot}$, see \S~\ref{sec:om}) supports the
latter conclusion, being also inconsistent with the
softest proposed equations of state for super-dense matter.
This plus the mass measurements of
PSR~J0437$-$4715 \cite{vbs+08} and PSR~J0621+1002 \cite{sna+02} confirm
beyond doubt the results of a previous statistical analysis of masses of
several binary MSPs in globular clusters \cite{fwbh08}, which showed
that MSPs have a much wider mass distribution than observed among the
members of double neutron star systems. The likely cause is the accretion
episode that spun up these NSs \cite{bv91,lor08}.

There are several reasons why \psr\, appears to be a normal MSP
(i.e., spun up by accretion of mass from a companion):
a) its spin period is typical of that of MSPs
and $\sim$8 times shorter than that of the fastest
young pulsar, PSR J0537$-$6910, near the N157B supernova remnant in the
Large Magellanic Cloud \cite{mgz+98} b) its magnetic field ($B_0$) is
similar to those observed among MSPs and two orders of magnitude
smaller than the smallest $B_0$ thus far observed for a young
pulsar, $3 \times 10^{10}$G for PSR~J1852+0040, a 105-ms X-ray pulsar near
the centre of the supernova remnant Kesteven~79 \cite{hg10} c)
its mass suggests it accreted mass from a former donor star.
Precise mass measurements for components of double neutron
star systems range from $1.25\,M_{\odot}$ \cite{ksm+06} to
$1.44\,M_{\odot}$ \cite{wnt10}; this shows that a large majority of NSs
form with masses near the Chandrasekhar limit \cite{lor08}.
If this was also the case for \psr\, then it must have gained
$\Delta M \sim 0.22-0.42\,M_{\odot}$ during accretion, which is
well within the range expected for low-mass and intermediate-mass binary
pulsars (Pfahl et al. 2002)\nocite{prp02}. All of this is inconsistent
with scenarios where \psr\, forms directly from the collapse of a single star.
Even alternative MSP formation scenarios (like merger of massive WDs
or accretion induced collapse) require
two stars to form the MSP.

\subsection{\psr\, is not a triple system}
\label{sec:notriple}
The discovery that the companion to \psr\, is the MS star
previously suspected of being associated with the system immediately
rules out the detailed triple system scenario proposed in
Champion et al. (2008). Our measurement of the variation of eccentricity
had already ruled out the possibility that the Kozai mechanism is the
origin of the observed eccentricity
(Gopakumar, Bagchi \& Ray 2009)\nocite{gbr09}.

Despite this, it is clear that the present companion
cannot be the star that recycled the pulsar. If it were, then it must have at
some point filled the region where matter will remain bound to it
(its ``Roche lobe'') in order to transfer matter to the companion.
The problem is that at the closest approach between the stars the
companion is $\sim 23$  times smaller than its present Roche lobe.
Being an unevolved MS
star it was even smaller in the past\footnote{If this star had ever
filled its Roche lobe the orbit would have been circularised. This is
inconsistent with the present large and non-varying eccentricity.}
and even less likely to have supplied the material that recycled this
pulsar.

Could the former donor have exchanged orbits with the main-sequence
companion and now orbit the \binary\,at a large distance? If there were
a third component in the system, the
\binary would be accelerating towards it, with a line-of-sight component
given by $a_{l}$. This should produce a variation of its orbital period
given by $\dot{P}_b/P_b = a_{l}/c$. Variations of $a_{l}$ due to the
relative motion of the \binary and the outer component should 
produce a variation of the first derivative of the spin frequency
($\dot{\nu}$), $\ddot{\nu}$
(e.g., Backer, Foster \& Sallmen, 1993)\nocite{bfs93}. Our measurements of
$\dot{P}_b$ and $\ddot{\nu}$ are not statistically significant
(see Table~\ref{tab:parameters}), i.e., we detect no third
component in this system. If the former donor still exists, it is no
longer bound to this system.

\subsection{Motion of \psr\, in the Galaxy}
\label{sec:motion}

\begin{figure*}
  \includegraphics[width=14cm]{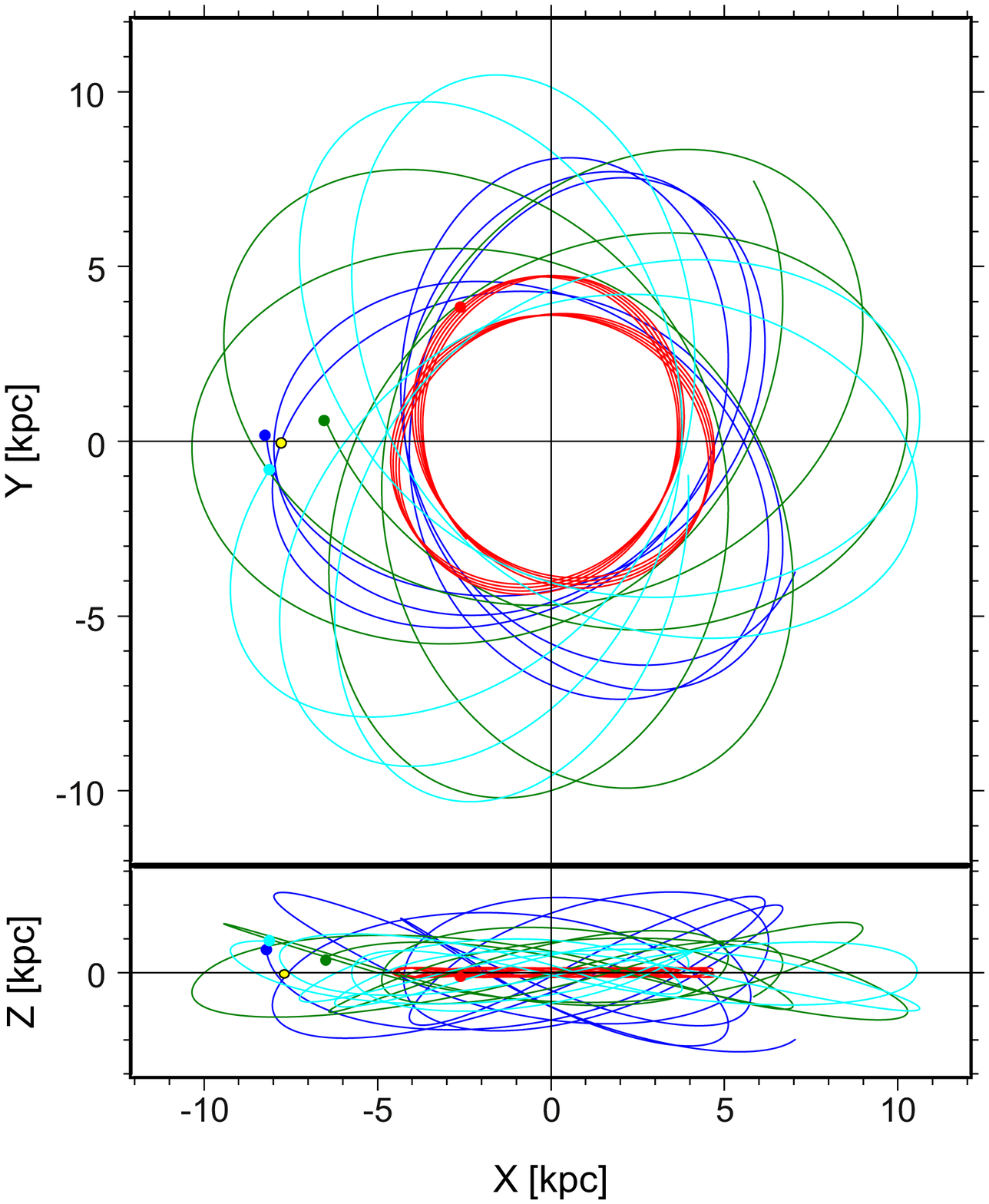}
  \caption{Present Galactic positions (dots) and past orbits,
    integrated for 1 Gyr (solid curves) of the four MSPs with known
    radial velocities: \psr\, (Red), PSR~J1012+5307 [Dark Blue,
      \citet{lwj+09}], PSR~J1023+0038 [Light Blue, \citet{ta05,asr+09}]
    and PSR~J1738+0333 [Olive Green]. The Sun's position is indicated
    by the yellow circle. In the top plot we look at the Galaxy from
    the North Galactic pole, in the bottom plot from the direction in
    the Galactic plane perpendicular to the line from the Sun to the
    Galactic centre. We can see that \psr\, is always near the
    Galactic plane, and that its orbital velocity about the Galaxy has
    a relatively small eccentricity.  }
  \label{fig:orbit}
\end{figure*}

A possible explanation for the absence of the former donor is that
it might have been exchanged by the present companion. Before the
discovery of \psr, this was the only mechanism known to produce MSP
binaries with eccentric orbits. This can only occur in environments
with high stellar densities such as the cores of globular clusters
\cite{phi92} or, hypothetically, the Galactic Centre. This is the
reason why all other eccentric binary MSPs are observed in globular
clusters \cite{rsb+04,fgri04,rhs+05,frb+08}. For this reason it was
proposed in \cite{crl+08} that \psr\, might have formed in such an
environment and subsequently have been ejected
by the recoil produced by the presumed exchange interaction.

To investigate this possibility, we used the radial velocity $\gamma$ and
the pulsar's Galactic coordinates, DM distance, and proper motion
to derive all three coordinates of its position in space and 
all three components of its velocity vector,
as in Wex et al. (2000)\nocite{wkk00} and Lazaridis et al.
(2009)\nocite{lwj+09}; the results are displayed in Table 1. We can thus
calculate the past trajectory of the binary in the Galactic potential
\cite{kg89}. 
We implement this process in a Monte Carlo simulation of 10000
orbits in the model Galactic potential. We take the uncertainties
of $\mu$, $\gamma$ and distance\footnote{We used a 1-$\sigma$ relative
uncertainty in the distance of 15\% such that it is similar to the range
of optically derived distances.} into account by integrating
many orbits with random initial conditions having a distribution
of parameters consistent with the observed values and uncertainties.
We also compare our results with those obtained using an alternative
model for the Galactic potential \cite{pac90}.

\begin{figure}
  \includegraphics[width=11.2cm]{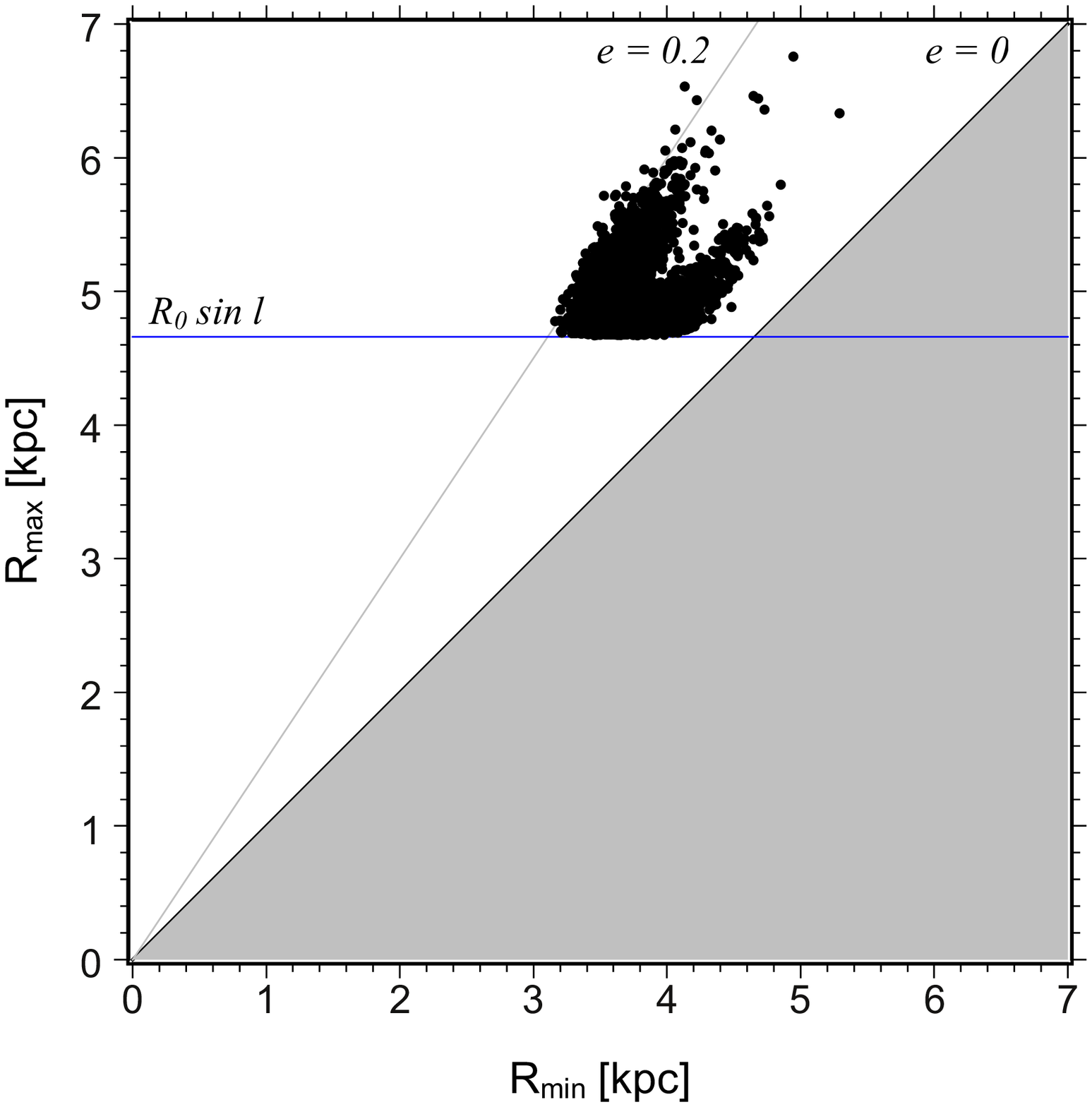}
  \includegraphics[width=10cm]{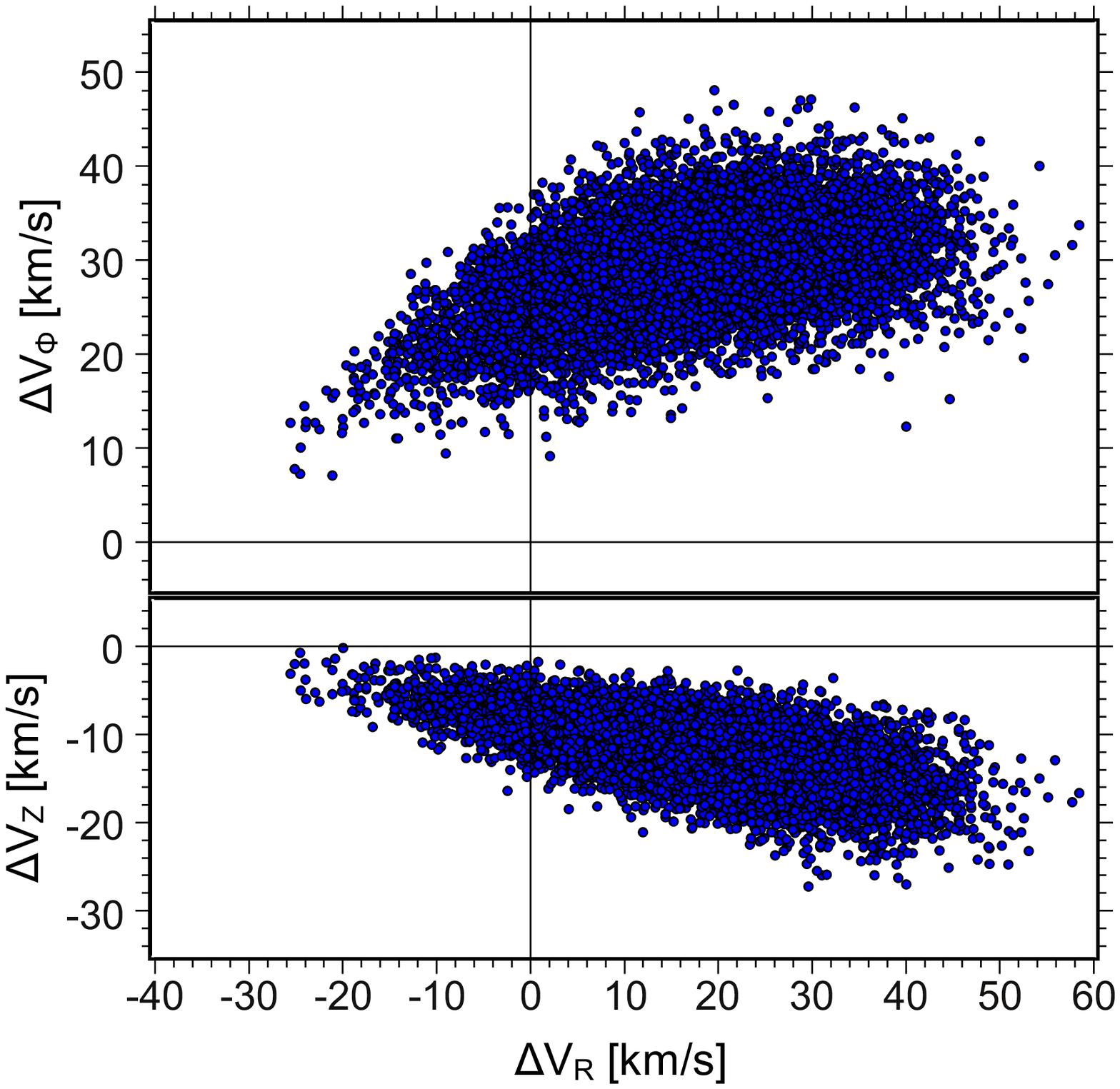}
  \caption{Monte-Carlo simulation of 10000 pulsar orbits. {\em Top:}
   Minimum and maximum distances from the centre of the Galaxy.
   We can see here that the pulsar is never within 3 kpc from the
   centre of the Galaxy
   {\em Bottom}: Starting velocities in previous simulation relative to
   each simulated pulsar's average of local velocities (i.e., the
   system's {\em peculiar velocity}).  }
  \label{fig:orbital_parameters}
\end{figure}

Our results are presented in Figs.~\ref{fig:orbit}
and \ref{fig:orbital_parameters}. The starting
orbits have a range of proper motions and systemic velocities
consistent with the uncertainties in Table 1.  We also assume a
1-$\sigma$ relative uncertainty of 15\% in the distance.  For each
point we integrate the pulsar orbit back in time for about 300 Myr and
record the maximum (R$_{\rm \max}$) and minimum (R$_{\rm \min}$)
distances to the centre of the Galaxy. The diagonal black line
shown in the top plot of Fig.~\ref{fig:orbital_parameters}
indicates circular orbits and the gray line eccentricities of
0.2. R$_{\rm max}$ cannot be smaller than the minimum possible
distance to the Galactic centre at present (blue line), this is given
by $R_0 \sin l$, where $R_0$ is the Sun's distance to the Galactic
centre (7.7 kpc in the model used for this simulation \cite{kg89})
and $l$ is the pulsar's Galactic longitude, $37.33^\circ$. 
In this plot we can see that the minimum possible distance from
the Galactic centre is always larger than 3 kpc. This
excludes formation in the Galactic centre and in the bulge
globular clusters.

It is also apparent from our simulations that
the distance from the Galactic plane never exceeds 270~pc (99.7\%
confidence limit), which is unusual for MSPs.
The globular clusters observed outside the bulge
have a wide distribution of Galactic heights \cite{har96}, which
implies that most of them orbit the Galaxy well outside the plane. The
formation of \psr\, in such a cluster would require that the system is
ejected at the exact time the cluster is crossing the Galactic plane
and that the resulting velocity is very nearly in the Galactic plane
and close to the average local Galactic rotation (the system's current
peculiar velocity is 37$\pm$9 km s$^{-1}$, see bottom plot of 
Fig.~\ref{fig:orbital_parameters}). We find this possibility highly
unlikely. The probability of formation in an exchange interaction
appears therefore to be extremely small.

\section{Formation of \psr}
\label{sec:formation}
To summarise, the measured mass of \psr\, (plus its small spin
period and magnetic field) make any scenarios where it forms as it
is unlikely; all pulsar characteristics are consistent with it
having been recycled (\S~\ref{sec:massimplications}).
Our limits on the variation of the eccentricity and our optical
detection exclude the triple system scenario proposed in
Champion et al. (2008, \S~\ref{sec:notriple}). In particular,
we find that the present companion is not the donor star that
recycled the pulsar; if this former donor still exists
it is no longer bound to the system. Furthermore, our calculations of the
3-D systemic motion (\S~\ref{sec:motion}) show that it is unlikely that
the system originated in an environment with high stellar density.
Therefore, it is unlikely that the former donor star was exchanged
by the present companion. So how did \psr\, form? And what happened to
the former donor star?

\subsection{Multiple star spiral-in}
\label{sec:multiple}
We suggest that the system started as a triple (indicated below
in square brackets), where the outer
companion (the present MS star companion) was initially on a much
wider orbit than at present. The inner binary system (indicated
in parentheses) had a much shorter
orbital period and consisted of two somewhat more massive MS stars:
\begin{math}
\\
\\
\rm [(MS + MS; 12\,+\,2 M_{\odot}; 20 d) +  MS; 1 M_{\odot}; 1000 d]
\\
\\
\end{math}
(the numbers are conjectural, we show them for illustrative purposes only).
The more massive star (the progenitor of the pulsar), evolves and becomes
a red supergiant (RSG):
\begin{math}
\\
\\
\rm [(RSG + MS; 12\,+\,2 M_{\odot}; 20 d) +  MS; 1 M_{\odot}; 1000 d]
\\
\\
\end{math}
At this point, it starts transferring mass to the other component in
the inner binary. In such a situation, mass transfer is generally
unstable, and the envelope of the RSG is expected to engulf the
companion and lead to a common envelope (CE) phase, where the helium
core of the RSG and the inner companion are embedded in the envelope
of the RSG. Because of friction with the envelope, the orbit of this
embedded binary will decay, releasing orbital energy in the
process. If this energy is sufficient to eject the envelope, this will
leave a much closer binary consisting of a helium star and the
more-or-less unaffected inner companion \citep[see, e.g.,][]{bv91}:
\begin{math}
\\
\\
\rm [(He + MS; 3\,+\,2 M_{\odot}; 3 d) +  MS; 1 M_{\odot}; 1000 d]
\\
\\
\end{math}
If the outer component of the triple system is close enough,
it may also be affected by this CE phase, since the envelope
will greatly expand because of the orbital energy that is being
deposited within it \citep[to radii of $\sim 5-10\,$AU; see, e.g.,][]{pod01}.
It will also be engulfed by the expanding envelope and will also experience a
spiral-in phase  \cite{ev86}. This will produce a much closer triple system.
Without realistic hydrodynamical simulations, it is difficult to determine
the post-CE parameters. Here, we will consider two potential outcomes,
which we will discuss in more detail later:
\begin{math}
\\
\\
\rm a)\,[(He + MS; 3\,+\,2 M_{\odot}; 2 d) +  MS; 1 M_{\odot}; 100 d]
\\
\\
\end{math}
or
\begin{math}
\\
\\
\rm b)\,[(He + MS; 3\,+\,2 M_{\odot}; 0.8 d) +  MS; 1 M_{\odot}; 50 d]
\\
\\
\end{math}
Eventually, the He core will explode in a supernova and produce a
NS. The sudden mass loss associated with this event will make
the orbits of the other two components wider and somewhat eccentric:
\begin{math}
\\
\\
\rm a)\,[(NS + MS; 1.4\,+\,2 M_{\odot}; 2.5 d) +  MS; 1 M_{\odot}; 120 d, e = 0.1]
\\
\\
\end{math}
or
\begin{math}
\\
\\
\rm b)\,[(NS + MS; 1.4\,+\,2 M_{\odot}; 8 hr) +  MS; 1 M_{\odot}; 70 d, e = 0.44]
\\
\\
\end{math}
This event could be either a normal Fe-core collapse supernova (SN) or
an electron-capture
(e-capture) supernova \citep[see][and further references therein]{plp+04}.
 Given the fact that the system 
remained bound after the explosion, it is not likely that the SN 
produced a major kick, nor large fractional mass loss, particularly
under scenario "a", where the system is wider and acquires
a small eccentricity. This is
consistent with the observed peculiar velocity for the system,
which is smaller than for the other systems where we can make
a complete determination of this quantity (see Fig.~\ref{fig:orbit}).
We note too that such SN characteristics are expected from 
e-capture supernovae.

\subsection{Post-supernova evolution}
At this stage the inner binary resembles the progenitor of a typical low- or
intermediate-mass X-ray binary (L/IMXB). Once the normal stellar
component of the inner binary has evolved to fill its Roche lobe, mass
transfer to the NS will start to increase its mass and spin
it up to millisecond periods (see, e.g., Pfahl et al. 2002)\nocite{prp02}.
Initially, when the separation of the inner binary is
sufficiently small compared to the periastron of the outer
component, the evolution of the L/IMXB will not be significantly
affected by the presence of the outer component.

If the mass-transfer rate is very high, mass loss from the inner
binary will cause the orbit of the outer component to widen to some
degree.  However, even for conservative mass transfer the inner binary
is generally expected to widen more drastically, in particular in the
2.5 day scenario (scenario ``a'' in \S~\ref{sec:multiple}; see
Podsiadlowski et al.\ 2002\nocite{prp02b}):
\begin{math}
\\
\\
\rm [(MSP + MS; 1.67\,+\,0.3 M_{\odot}; 50 d) +  MS; 1 M_{\odot}; 150 d, e = 0.1]
\\
\\
\end{math}
until its separation reaches a critical fraction of the periastron distance of
the outer companion. This leads to a chaotic three-body interaction
\citep[see, e.g.,][]{hut84,phi93a}. One of the most probable outcomes is the
ejection of the least massive component -- in this case most likely
the mass donor in the inner binary, leaving a bound, eccentric binary system
consisting of the MSP and the formerly outer companion, in an orbit tighter
than before:
\begin{math}
\\
\\
\rm (MSP + MS; 1.67\,+\, 1 M_{\odot}; 95 d, e = 0.44)
\\
\\
\end{math}
If the starting orbital period is small ($P_B < 8$ hr, scenario b in
\S~\ref{sec:multiple}), then gravitational radiation might lead to
the destruction of the remnant of the former donor by the newly formed MSP,
very much in the same way as PSR 1957+20, the black-widow pulsar
(Fruchter et al. 1988)\nocite{fst88}, is evaporating its companion
(although in the latter case the
process might take several Hubble times). This is a possible
formation channel for isolated MSPs. Since these represent
$\sim$20\% of the known MSPs in the Galactic disk, this should not be
too unlikely a scenario. The end result would also resemble the \binary.

\subsection{Alternative scenarios}
\label{sec:alternative}
One can imagine alternative triple scenarios to the one outlined
above. For example, it is possible that the progenitor of the MSP
was initially the outer component of a stable triple system.
When this massive star evolved to become a red supergiant, it could have
engulfed the inner binary completely, leading to a spiral-in of that
binary inside the common envelope. As it spirals in, its orbit
could become disrupted by the tide induced by the more massive
component; both of its components then start orbiting the core of
the more massive He star independently, but at different distances
given their different orbital velocities prior to disruption.
This would produce a result quite similar to the double star
spiral-in. Another possibility is that this binary survived the
CE phase, but it was later disrupted when its orbital radius increases
due to a mass transfer stage.

\section{Conclusions}\label{sec:conclusions}
Our Arecibo and Green Bank
radio timing observations include a full determination of the
relativistic Shapiro delay, a very precise measurement of
the apsidal motion and new constraints of the orbital orientation
of the system. Through a detailed analysis of all of these,
we derive new constraints on the masses of the pulsar and
companion. The mass of the pulsar ($1.667 \pm 0.021$~M$_{\odot}$)
adds to a growing population of NSs with masses which
significantly exceed 1.4~M$_{\odot}$ and rules out some of the
proposed equations of state of superdense matter. This high mass
also suggests that the pulsar was recycled.

Our observations with the Very Large Telescope reveal
shifts in the spectral lines of the suspected companion to
\psr, which show that it has the 95-day orbital motion
expected for the pulsar's binary companion.
This star has characteristics very similar
to the Sun; making this system unique among all known binary MSPs.
A detailed discussion of the results of the radio timing and
optical spectroscopy rules out most proposed formation mechanisms.
We are then forced to accept the only remaining possibility, i.e., that the
system originated from a triple system with a compact inner component
and the presently observed MS companion as the outer component.
The inner compact binary then formed the observed MSP, with the
elimination of the former donor star (i.e., evolution towards an
``isolated'' MSP), or its ejection from the system.

Detailed stellar evolution simulations of the formation of
this system might provide new insights on how MSPs form. As an example, the
mass loss and kick velocities associated with the supernova event that
formed this NS must have been small enough to allow the whole
system to remain bound. Such a gentle formation would produce a small
recoil velocity for the whole system; this is consistent with the
small peculiar velocity observed at present.  This would generally
favour alternative small-kick formation mechanisms, like electron
capture supernovae. This means that a dynamical study of this system
could lead to a conclusive demonstration of alternative formation
channels for NSs in general. Some of these exotic mechanisms
(for example, white dwarf mergers) could be especially important for
explaining the existence of isolated MSPs.

\section*{Acknowledgements}
  This work was supported by the
  NSF through a cooperative agreement with Cornell University to
  operate the Arecibo Observatory. The National Radio Astronomy
  Observatory is a facility of the National Science Foundation
  operated under cooperative agreement by Associated Universities,
  Inc. Pulsar research at UBC is supported by an NSERC Discovery Grant.
  V.M.K. receives support via an NSERC Discovery Grant, CIFAR,
  FQRNT, and from the Lorne Trottier Chair and a Canada Research Chair.
  J.W.T.H. is a Veni Fellow of The Netherlands Organisation for
  Scientific Research (NWO). M.A.M and D.R.L are supported by the
  West Virginia EPSCoR program and the Research Corporation for
  Scientific Advancement. J.P.W.V is supported by the European Union
  under Marie Curie Intra-European Fellowship 236394. D.J.N. was
  supported by NSF grant AST 0647820 to Bryn Mawr College. We also
  thank the referee, Peter P. Eggleton, for the careful review and
  many constructive suggestions.

\bibliographystyle{mn2e}
\bibliography{}

\end{document}